\documentclass[longauth]{aa} 
\usepackage{graphicx}
\usepackage{txfonts}

\usepackage{ulem}
\usepackage{natbib,twoopt}
\usepackage{xcolor}
\usepackage{aecompl}
\usepackage{supertabular,booktabs}
\usepackage{float}
\usepackage{amssymb}
\usepackage{hyperref}
\usepackage{url} 
\usepackage{rotating}

\hypersetup{
        colorlinks = True, 
     linkcolor = {blue},
     linkbordercolor = {blue},
     anchorcolor=green,
     citecolor= {blue} }

\newcommand{\msun}{M$_\odot$}
\newcommand{\alpfe}{~[$\alpha$/Fe]~}
\newcommand{\parsec}{~\texttt{PARSEC}~}
\newcommand{\yfe}{~[Y/Fe]~}
\newcommand{\bafe}{~[Ba/Fe]~}
\newcommand{\ali}{A$_\mathrm{Li}$~}
\newcommand{\alinlte}{A$_\mathrm{Li}^{NLTE}$~}
\newcommand{\teff}{T$_\mathrm{eff}$~}
\newcommand{\logg}{$\log(g)$~}
\newcommand{\lialpha}{Li-$[\alpha/Fe]$ anticorrelation~}
\newcommand{\feh}{[Fe/H]~}
\newcommand{\mh}{[M/H]~}

\begin{document}

   \title{The $Gaia$-ESO Survey: Lithium enrichment histories \\of the Galactic thick and thin disc}

   \author{X. Fu \inst{1,2},
          D. Romano \inst{2},
          A. Bragaglia \inst{2},
          A. Mucciarelli \inst{1,2},
          K. Lind \inst{3},
          E. Delgado Mena \inst{4},
          S. G. Sousa \inst{4},
          S. Randich \inst{5},
          A. Bressan \inst{6},
          L. Sbordone \inst{7},
          S. Martell \inst{8},
          A.~J. Korn \inst{9},
          C. Abia \inst{10},
          R. Smiljanic \inst{11},
          P. Jofr\'e  \inst{12},
          E. Pancino \inst{5},
          G. Tautvai\v{s}ien\.{e} \inst{13},
          B. Tang \inst{14},
          L. Magrini \inst{5},
          A.~C. Lanzafame \inst{15},
          G. Carraro \inst{16},
          T. Bensby \inst{17},
          F. Damiani \inst{18},
          E.~J. Alfaro \inst{19},
          E. Flaccomio \inst{18},
          L. Morbidelli \inst{5},
          S. Zaggia \inst{20},
          C. Lardo \inst{21},
          L. Monaco \inst{22},
          A. Frasca \inst{15},
          P. Donati \inst{2},
          A. Drazdauskas \inst{13},
          Y. Chorniy \inst{13},
          A. Bayo \inst{23},
          G. Kordopatis \inst{24}
          }

   \institute{
           Dipartimento di Fisica \& Astronomia, Universit\`{a} degli Studi di Bologna,
             via Gobetti 93/2, 40129 Bologna, Italy 
         \and 
             INAF - Osservatorio Astronomico di Bologna,
             via Gobetti 93/3, 40129 Bologna, Italy  
             \email{xiaoting.fu@oabo.inaf.it}
         \and
             Max-Planck Institut f\"{u}r Astronomie, 
             K\"{o}nigstuhl 17, 69117 Heidelberg, Germany 
         \and
         Instituto de Astrof\'{\i}sica e C\^{e}ncias do Espa\c{c}o, 
         Universidade do Porto, CAUP, Rua das Estrelas, PT4150-762 Porto, Portugal 
         \and
         INAF - Osservatorio Astrofisico di Arcetri, 
         Largo E. Fermi 5, 50125, Florence, Italy 
         \and
          SISSA - International School for Advanced Studies, 
          via Bonomea 265, 34136 Trieste, Italy 
         \and
         European Southern Observatory, 
         Alonso de Cordova 3107 Vitacura, Santiago de Chile, Chile 
         \and
         University of New South Wales Sydney, Australia 2052 
          \and
         Department of Physics and Astronomy, 
         Uppsala University, Box 516, SE-751 20 Uppsala, Sweden  
          \and
         Dpto. F\'{\i}sica Te\'{\o}rica y del Cosmos, 
         Universidad de Granada, 18071 Granada, Spain 
         \and
          Nicolaus Copernicus Astronomical Center, Polish Academy of Sciences, ul. 
         Bartycka 18, 00-716, Warsaw, Poland 
         \and
         Institute of Astronomy, University of Cambridge,
         Madingley Road, Cambridge CB3 0HA, United Kingdom  
         \and
         Institute of Theoretical Physics and Astronomy, Vilnius University,
         Saul\.{e}tekio av. 3, LT-10257 Vilnius, Lithuania 
         \and
         Departamento de Astronomia,
         Universidad de Concepcion,
         Concepcion, 3349001, Chile 
         \and
         Dipartimento di Fisica e Astronomia, 
         Sezione Astrofisica, Universit\'{a} di Catania, 
         via S. Sofia 78, 95123, Catania, Italy 
         \and
         Dipartimento di Fisica e Astronomia,  Universit\`{a} di Padova,
         Vicolo Osservatorio 3, 35122 Padova, Italy 
         \and
         Lund Observatory, Department of Astronomy and Theoretical Physics, 
         Box 43, SE-221 00 Lund, Sweden  
         \and
         INAF - Osservatorio Astronomico di Palermo, 
         Piazza del Parlamento 1, 90134, Palermo, Italy  
         \and
         Instituto de Astrof\'{i}sica de Andaluc\'{i}a-CSIC, Apdo. 
         3004, 18080, Granada, Spain 
         \and
         INAF - Padova Observatory, 
         Vicolo dell'Osservatorio 5, 35122 Padova, Italy  
         \and
         Laboratoire d'astrophysique, Ecole Polytechnique F\'ed\'erale de Lausanne (EPFL), 
         Observatoire de Sauverny, CH-1290 Versoix, Switzerland   
         \and
          Departamento de Ciencias Fisicas, Universidad Andres Bello, 
         Fernandez Concha 700, Las Condes, Santiago, Chile 
         \and 
         Instituto de F\'{i}sica y Astronomi\'{i}a, 
         Universidad de Valparai\'{i}so, Chile  
         \and
         Observatoire de la C\^{o}te d'Azur,
         Laboratoire Lagrange, CNRS UMR 7293,
         Nice Cedex 04, France 
                                       }

   \date{Received --, 2017; accepted 9 November, 2017}

  \abstract
   { Lithium abundance in most of the warm metal-poor main sequence stars shows
   a constant plateau  (A(Li) $\sim$ 2.2 dex) and then the upper envelope of
   the lithium vs. metallicity distribution increases as we approach solar
   metallicity.  Meteorites, which carry information about the chemical
   composition of the interstellar medium at the solar system formation time,
   show a lithium abundance A(Li)$\sim$ 3.26 dex.  This pattern reflects the Li
   enrichment history of the interstellar medium during the Galaxy lifetime.
   After the initial Li production in Big Bang Nucleosynthesis, the sources of
   the enrichment include AGB stars, low-mass red giants, novae, type II
   supernovae, and Galactic cosmic rays.  The total amount of enriched Li is
   sensitive to the relative contribution of these sources.  Thus different Li
   enrichment histories are expected in the Galactic thick and thin disc.  We
   investigate the main sequence stars observed with UVES in $Gaia$-ESO Survey
   iDR4 catalog and find a Li-\alpfe anticorrelation independent of \feh,
   \teff, and \logg.  Since in stellar evolution different $\alpha$
   enhancements at the same metallicity do not lead to a measurable Li
   abundance change, the anticorrelation indicates that more Li is produced
   during the Galactic thin disc phase than during the Galactic thick disc
   phase.  We also find a correlation between the abundance of Li and
   s-process elements Ba and Y, and they both decrease above the solar
   metallicity, which can be explained in the framework of the adopted
   Galactic chemical evolution models.  }

   \keywords{stars:  abundances --   
            Galaxy: abundances --  
            Galaxy: disc }   
  \authorrunning{Fu, X. et al.}
  \titlerunning{Li enrichment histories of the Galactic thick and thin disc}
  \maketitle

\section{Introduction}
\label{intro}

  The evolution of lithium ($^7$Li) in the Galaxy is a critical and not well
  understood issue, because of the many unknowns that still affect the proposed
  production and destruction channels of this element.  Most of the metal-poor
  ($-$2.4~$\lesssim$~[Fe/H]~$\lesssim -$1.4), warm (5700~K~$\lesssim T_\mathrm{eff} \lesssim$~6800~K) 
  Galactic halo dwarfs are known to share a
  very similar $^7$Li abundance, the so-called \textit{"Spite plateau"}
  \citep[A(Li)\footnote{A(Li) =$ 12 + \log [n({\rm Li})/n({\rm H})]$ where $n$
  is the number density of atoms and 12 is the solar hydrogen abundance.}
  $\simeq$~2.05\,--\,2.2 dex,][]{spite82, bonifacio97, Asplund2006,
  bonifacio07}, which for a long time has been thought to represent the
  primordial Li abundance produced in the Big Bang Nucleosynthesis.  However,
  Standard Big Bang Nucleosynthesis (SBBN) predicts that the primordial Li
  abundance is mainly determined by the primordial baryon-to-photon ratio which
  can be derived from the acoustic oscillations of the cosmic microwave
  background (CMB), and the baryon-to-photon ratio obtained by the Planck
  satellite data leads to  A(Li)~=~2.66\,--\,2.73 dex \citep{coc14}, i.e. the
  SBBN predicted primordial Li abundance exceeds the stellar observationally
  inferred value by a factor of almost three.  Furthermore, the Spite plateau
  is not constant, but bends down for extremely metal-poor stars \citep[\feh
  $\lessapprox-$2.8 dex, e.g. ][]{sbordone10, Melendez2010,hansen14,
  Bonifacio15}.  The environmental $^7$Li evolution model proposed by
  \citet[][]{fu15} offers a way to reconcile the SBBN theory to the
  observations of stars on the plateau by considering the pre-main sequence
  accretion, overshooting, and main sequence diffusion, while also accounting
  for the drop of $^7$Li abundances in extremely metal-poor stars, but it needs
  to be studied more and better constrained (for instance, by including the
  effects of stellar rotation).

  As far as the Galactic disc stars are concerned, a large dispersion is seen
  in the data \citep[see][for recent
  works]{ramirez12,delgadomena15,guiglion16}.  The observed scatter is due to
  efficient $^7$Li destruction in stellar interiors \citep[e.g.,
  A$_\mathrm{Li}$~=~1.05 dex in the Sun, ][]{grevesse07}, coupled to
  non-negligible $^7$Li production on a Galactic scale. The discovery of
  unevolved stars with high $^7$Li abundances \citep[comparable or higher than
  those observed in meteorites, A(Li)~=~3.26 dex, ][]{lodders09}, dates to long
  ago \citep[see e.g.][for T\,Tauri, F and G dwarf, and Hyades main-sequence
  stars, respectively]{bonsack60,herbig65,wallerstein65}, and it is now
  commonly interpreted as a signature of $^7$Li enrichment of the interstellar
  medium (ISM), due to various production processes. Indeed, $^7$Li is
  synthesized in different astrophysical sites, apart from the Big Bang:
  
  \begin{itemize}
          \item High-energy processes involving Galactic cosmic rays (GCRs) hitting the 
                  ISM atoms contribute to less than 20\,--\,30 per cent to 
                  the meteoritic $^7$Li abundance \citep[][]{reeves70,meneguzzi71,lemoine98,
                  romano01,prantzos12}, so that
                  about 70\% of the Solar System $^7$Li must 
                  originate from thermonuclear reactions acting in the stellar interiors.
          \item As first suggested by \citet[][]{domogatskii78}, the $\nu$-process may 
                  lead to the synthesis of $^7$Li in the helium shell of core-collapse supernovae 
                  (SNe). However, this mechanism ought to work quite inefficiently at a Galactic 
                  level \citep[see discussions in][]{vangioniflam96,romano99}; furthermore, to 
                  the best of our knowledge there is no observational evidence in support of it. 
          \item Observations in both 
                  the Milky Way \citep[disc, bulge, halo, Globular Clusters,
                  e.g.] []{wallerstein82,pilachowski86,hill99,kraft99,
                  balachandran00,kumar09,monaco11,DOrazi2015,
                  casey16,kirby2016} and Local Group dwarf galaxies
                  \citep{dominguez2004,kirby2012} have conclusively shown that
                  a  number of Li-rich giants do exist with abundances higher
                  than the standard stellar evolution theory predicts
                  (A(Li)$\sim$ 1.0 dex at the beginning of the Red Giant Branch
                  bump for Pop. II stars around 1 \msun \citep{Charbonnel2007,
                  fu2016}, and  even lower for more massive stars
                  \citep{Iben1967a, Iben1967b} ), or even exceeding the SBBN
                  prediction and  the meteorites' value, indicating that the Li
                  in these stars must have been created rather than preserved
                  from destruction.  Stars can produce $^7$Li in their late
                  stages of evolution via the Cameron-Fowler mechanism
                  \citep[][]{cameron71}, which may work both in
                  intermediate-mass stars on the asymptotic giant branch (AGB)
                  \citep[][]{sackmann92} and in low-mass stars on the red giant
                  branch (RGB), under special conditions
                  \citep[][]{sackmann99}.  \citet{Abia1993} conclude that the
                  contribution of carbon stars can reach 30\% to the Galactic
                  Li enrichment depending on the mass loss and Li production in
                  the Li-rich AGB stars.  However, the detailed mechanism(s) of
                  the Li enrichment in these giant stars is still an open
                  problem, and we lack a clear and unequivocal assessment of
                  the relative contributions of these two stellar contributions
                  to the overall Galactic lithium enrichment: for instance,
                  \citet[][]{romano01} and \citet[][]{travaglio01} reach
                  opposite conclusions, because of the adoption of different
                  yield sets and different assumptions about the underlying
                  stellar physics (e.g. mixing process, mass loss rate) in
                  their chemical evolution models.
          \item Another, potentially major source of $^7$Li are nova systems, which are 
                  able to produce it when a thermonuclear runaway occurs in the
                  hydrogen envelope of the accreting white dwarf, as proposed
                  long ago by \citet[][]{starrfield78}. Notwithstanding the
                  considerable observational efforts, a direct detection of
                  $^7$Li during a nova outburst remained elusive until the very
                  recent detection of the blueshifted Li\,{\small
                  I}~$\lambda$\,6708 \AA~line in the spectrum of the nova
                  V1369\,Cen by \citet[][]{izzo15}. Based on the intensity of
                  the absorption line and on current estimates of the Galactic
                  nova rate, \citet[][]{izzo15} estimate that novae might be
                  able to explain most of the enriched lithium observed in the
                  young stellar populations. Further support in favour of a
                  high production of $^7$Li during nova outbursts comes from
                  the detection of highly enriched $^7$Be (later decaying to
                  $^7$Li) in the ejecta of novae V339\,Del, V2944\,Oph, and
                  V5668\,Sgr by \citet[][]{tajitsu15} and
                  \citet[][]{tajitsu16}, respectively \citep[see
                  also][]{molaro16}.
  \end{itemize}

  In this paper, we investigate the lithium content and  enrichment history in
  (mostly) thin and thick disc stars in the Milky Way, as well as its
  relationships with the abundances of selected $\alpha$ and $s$-process
  elements. To do so, we take advantage of UVES spectra analysed by the
  $Gaia$-ESO consortium for 1399 main sequence stars. Furthermore, we use an
  updated version of the model presented by \citet[][]{romano99,romano01}, that
  takes all of the above-mentioned sources of lithium into account \citep[see
  also][]{izzo15}, to discuss our $^7$Li measurements. In particular, we show
  that, in principle, in the framework of such a model it is  possible to offer
  an explanation for the observed peculiar trend of decreasing $^7$Li for
  super-solar metallicity stars \citep[][]{delgadomena15,guiglion16}, if it is
  real.

  The paper is organised as follows.  In section ~\ref{sec:ges} we describe the
  observations, data analysis, and sample selection.  The main results are
  presented in section ~\ref{sec:res}, including the trends of Li abundances in
  thin and thick disc stars, also in relation with the $\alpha$ and s-process
  elements behaviours.  In section~\ref{sec:dis} we discuss the Galactic
  chemical evolution of Li.  The conclusions and a final summary are given in
  section ~\ref{sec:con}.

\section{$Gaia$-ESO Survey observations and data analysis}
\label{sec:ges}

  The $Gaia$-ESO Survey \citep[GES hereafter,][] {gilmore12,randich13} is a
  large, public, high-resolution spectroscopic survey using the FLAMES facility
  at ESO Very Large Telescope (VLT), i.e. it simultaneously employs the UVES
  and GIRAFFE spectrographs \citep{dekker00,pasquini00}. GES is intended to
  complement the $Gaia$ astrometric and photometric data of exquisite precision
  \citep{gaia2016} with radial velocities and detailed chemistry for about
  $10^5$ stars of all major Galactic populations and about 70 open clusters
  (which will not be considered further in the paper since the purpose of this
  paper is to investigate the chemical evolution of the field stars.)

  To study the Milky Way (MW) field population, GES uses mostly the
  medium-resolution spectrograph GIRAFFE with two setups (HR10, HR21).
  However, GES also observes MW field stars with UVES and a setup centred at
  580nm ($\lambda=480-680$ nm, $R\sim47,000$) and is collecting
  high-resolution, high signal-to-noise (SNR) spectra of a few thousand FG-type
  stars in the solar neighbourhood (within about 2 kpc from the Sun). This
  sample includes mainly disc stars of all ages and metallicities, plus a small
  fraction of local halo stars, and it is intended to produce a detailed
  chemical and kinematical characterization, to fully complement the results of
  the $Gaia$ mission, which achieve their maximum precision within this
  distance limit.

  The detailed description of the MW sample selection is presented in
  \citet{stonkute16}.  In brief, the UVES observations are made in parallel
  with the GIRAFFE ones and their position and exposure time are dictated by
  the latter.  The 2MASS infrared photometry \citep{2mass} was used to select
  dwarfs and main sequence turn-off stars of FG spectral type, with a limiting
  magnitude of $J_{2MASS}=14$ and a narrow colour range.  The selection takes
  into account also the effects of interstellar extinction, estimated through
  the reddening maps of \citet{sfd98} Though some remaining uncertainties still
  affect the evolutionary phases of the stars, the majority of the observed
  targets turn out to be dwarf stars, as we will show later in
  Sec.~\ref{subsec:sample}.

  Multiple analysis pipelines are used by GES to produce stellar parameters and
  chemical abundances. The working group (WG) that takes care of FGK-type UVES
  spectra is WG11, split in different nodes producing independent results
  \citep{smiljanic14}.  The UVES spectrum reduction is centralized
  \citep[see][for a description]{sacco14} and each node is provided with 1-d,
  wavelength calibrated, and sky subtracted spectra.  Up to 11 nodes produced
  UVES results for iDR4, they have in common the stellar atmospheres \citep[1-d
  MARCS,][]{marcs}, the line list \citep{ruffoni14,heiter15}, and the solar
  reference abundances \citep{grevesse07}, see \citet{Pancino2017} for the
  calibration strategy.  The node results are homogenized by WG11 and
  subsequently by WG15 \citep[see for instance][for more details]{casey16},
  producing the recommended parameters and chemical abundances released first
  internally and then to the public.  We use here the internal data release 4
  (iDR4) which contains 54,490 stars observed up to the end of June 2014 (i.e.,
  30 months); the public version, released after a further validation process,
  can be accessed through the ESO Phase 3 webpage (http://www.eso.org/qi/).

  \subsection{Sample selection} \label{subsec:sample}

\begin{figure}
            \centering
            \includegraphics[width=0.5\textwidth,angle=0]{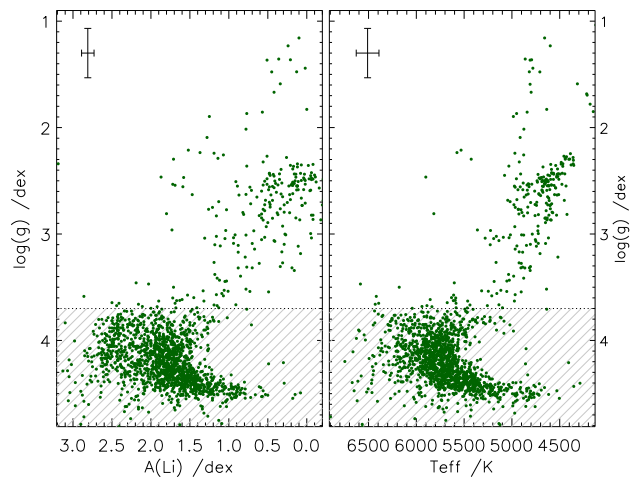}
            \caption{Field stars in GES iDR4 catalog with UVES observations.
               The left panel displays the Li content (LTE) against  \logg, and
               the right panel shows \logg, as an indicator of the evolutionary
               phase, against effective temperature (\teff).  We consider stars
               with \logg $\ge 3.7$ dex, namely those falling in the shaded
               areas, as main sequence stars.  The median values of the
               parameter error are displayed in the left upper corner of each
               panel.  }
            \label{fig:alilogg}
    \end{figure}

    To investigate the Galactic lithium enrichment history we select
    well-measured main sequence field stars with UVES spectra from the GES iDR4
    catalog.  In our selection 1884 UVES stars are marked as field stars,
    including those of the Galactic disc and halo designated as Milky Way
    ($GE\_MW$) fields, standard CoRoT ($GE\_SD\_CR$) field, standard radial
    velocity ($GE\_SD\_RV$) field, and stars to the Galactic Bulge direction
    ($GE\_MW\_BL$).  Note that these classifications are also used in the
    public GES data releases.  We set the surface gravity \logg=3.7 dex as the
    criterion to separate the dwarf and giant stars, Fig. \ref{fig:alilogg}
    illustrates the division.  The dwarfs largely outnumber the interloping
    giants (in our field UVES sample we have 1524 dwarfs and 360 giants,
    respectively), this indicates that the GES target selection was successful.

     We exclude the evolved giant stars in this study because their Li
     abundances are altered by internal mixing processes and do not reflect
     anymore the Galactic Li enrichment history.  When stars experience their
     first dredge-up after leaving the main sequence the surface convective
     zone deepens and brings materials from the hot interior to the stellar
     surface.  Lithium, which is easily destroyed at several million Kelvin,
     has its abundance significantly decreased  at this stage because of both
     dilution and destruction.  In fact, we also use this Li abundance drop to
     ensure that our dwarf-giant separation at \logg=3.7 is reliable (see the
     left panel of Fig.~\ref{fig:alilogg}).  We note that there are several
     giants in this figure that do not follow the A(Li)-\logg trend, these
     stars have already been discussed by \citet{casey16} who report on the
     "Li-rich giant problem" using GES data.  Compared to the giant stars, main
     sequence stars with the same stellar mass have a much thinner surface
     convective zone that could save Li from destruction.  In this paper we
     focus on these main sequence stars with \logg $\ge$ 3.7 dex. 
   
\begin{figure}
             \centering
             \includegraphics[width=0.5\textwidth,angle=0]{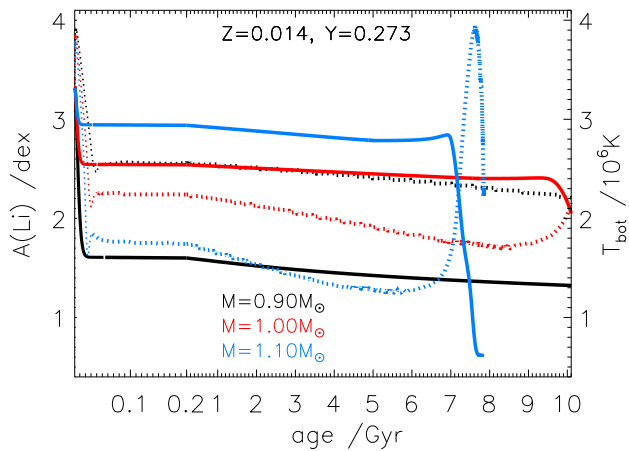}
             \caption{Pure effect of microscopic diffusion on Li abundance during the first 10~Gyr from stellar model \parsec.
                     Neither envelope overshooting nor pre-main sequence accretion is applied.
                     Li abundances (solid lines) and 
                     the temperature at the bottom of surface convective zone ($T_{bot}$, dotted lines)
                     are displayed as a function of stellar age.
                     Different colors indicate different stellar mass as indicated in the legend.  }
             \label{fig:diffu}
     \end{figure}

    However, for main sequence star another long-term stellar process,
    microscopic diffusion, could also lead to a depletion of the surface
    elements as the star ages \citep[see e.g.][]{richard02, Korn2007,
    Xiong2009, Nordlander2012, fu15, Dotter2017}.  To ensure that different
    ages of our sample stars do not introduce a significant dispersion in Li
    abundance, we check the pure effect of microscopic diffusion using the
    stellar model code \parsec \citep[V1.2s,][]{bressan2012}.  Three kinds of
    microscopic diffusion, including pressure diffusion, temperature diffusion,
    and concentration diffusion, as described in \citet{Thoul1994}, are
    considered.  To simulate the pure effect of microscopic diffusion, neither
    extra mixing (e.g. envelope overshooting) nor pre-main sequence accretion
    is applied in the models.  In Fig.~\ref{fig:diffu} three solar metallicity
    stars with different masses are shown as examples.  Starting from the
    meteorites' Li abundance, the main Li destruction in these stars occurs
    early in stellar life \citep[see also ][]{Chen2001}.  During their main
    sequence phase Li abundances (solid lines) decrease $\sim$0.2 dex.  Their
    temperatures at the bottom of the surface convective zone ($T_{bot}$,
    dotted lines), which is an indicator of the rate of Li burning,  are so low
    (1.3$\sim$2.5 $\times 10^6$ K) that even 7-10~Gyr of evolution can not lead
    to a notable destruction.  Thus Li depletion from nuclear reaction is
    negligible  in these stars during the  main sequence phase, the main
    depletion comes from the microscopic diffusion.  Microscopic diffusion
    takes 7-10~Gyr to decrease $\sim$ 0.2 dex of the surface Li abundance,
    which is the typical Li measurement uncertainty of our sample.  If extra
    mixing, which  slows the diffusion \citep{Xiong2009}, is considered, the
    depletion time will be even longer.  Therefore,  even a 7-10 Gyr age
    difference of the Galactic field stars will not affect our investigation.
    For stars with lower mass thus deeper surface convective zone (e.g.
    M=0.90\msun~ star in Fig.~\ref{fig:diffu}), their Li depletion can be
    traced by \teff, we will discuss it in detail in Sec. ~\ref{sec:res}.
    
    \begin{figure}
            \centering
            \includegraphics[width=0.5\textwidth,angle=0]{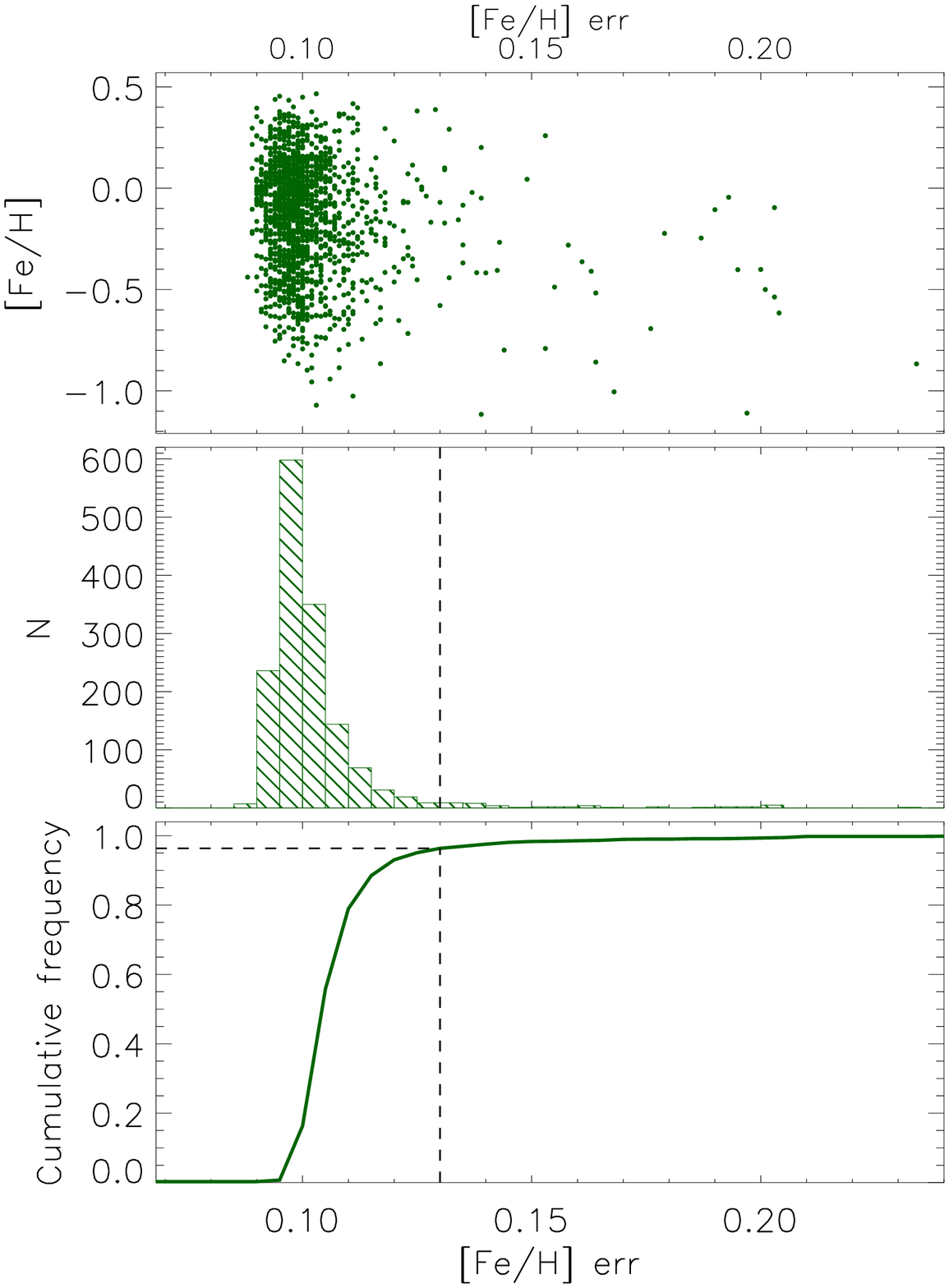}
            \caption{Distribution of [Fe/H] error for main sequence field UVES stars in GES iDR4 catalog.
              The upper panel displays the relation of [Fe/H] and [Fe/H] error, 
              while the center and lower panel show the histogram and cumulative frequency of [Fe/H] error, respectively.
              Ninety-five percent of stars lie in the interval smaller than 0.13 dex (vertical dashed lines in the last two panels). }
            \label{fig:fehhist}
    \end{figure}

    We then exclude  stars with large [Fe/H] errors.  Fig.~\ref{fig:fehhist}
    shows the distribution of [Fe/H] error for the 1524 main sequence field
    UVES stars; those with [Fe/H] error < 0.13 dex (95\% of the sample) are
    selected as "well-measured" objects.  Some of the well-measured stars are
    members of multiple stellar system, or have emission lines in their
    spectra.  Since multiple stellar system members carve up their initial Li
    \citep[e.g. in binary stars the primary one has higher Li abundance,]
    []{GonzalezHernandez2008, Aoki2012, fu15} and the emission lines are likely
    from pre-main sequence stars which may undergo a Li-depletion, we remove
    these stars from our sample to avoid confusion.  This reduces the sample to
    1432 objects.  In GES iDR4, 1097 of them are labeled as stars with Li upper
    limits, and 335 as true measurements.  We double check the spectra of stars
    with labeled Li measurements and mark 302 of them as high quality spectra
    around the Li resonance line.  In our final sample we thus have 1399
    well-measured UVES main sequence field stars sorted into two categories:
    \textit{i)} labeled Li upper limits (1097 stars, with signal-to-noise ratio
    (SNR) ranging in 12 - 259 and median SNR=63), and \textit{ii)} checked Li
    measurements (302 stars, SNR from 18 to 319, and median SNR=83).

   \section{Results}
   \label{sec:res}

\begin{figure*}
           \centering
           \includegraphics[width=0.95\textwidth,angle=0]{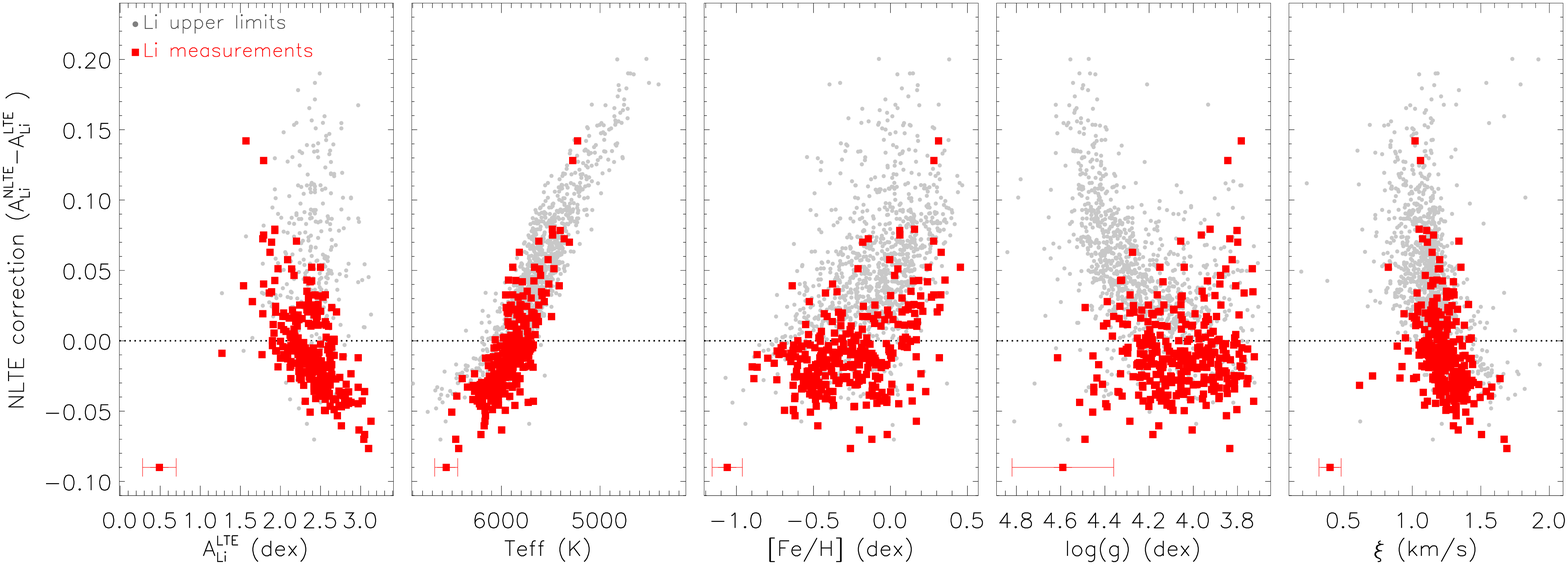}
           \caption{NLTE Li abundances corrections verses various parameters.
            The filled red squares are stars with checked Li measurements (302 stars), 
           and grey dots represent those labeled with Li upper limits in GES iDR4 catalog (1097 stars).
            The median error of each parameter is displayed in the left lower corner of each panel.}
           \label{fig:nlte}
   \end{figure*}

 The abundance analysis of Li in GES iDR4 is based on the local thermodynamic
 equilibrium (LTE) assumption, however in reality the line formation in stellar
 atmosphere is a more complex process, and non-local thermodynamic equilibrium
 (NLTE) effect should be accounted for.  We apply the NLTE corrections to the
 Li I $\lambda=6708$ \AA ~line using the correction grid from \citet{Lind2009}.
 Fig.~\ref{fig:nlte} displays the correction departures
 ($A_{Li}^{NLTE}$-$A_{Li}^{LTE}$) as a function of each parameter
 ($A_{Li}^{LTE}$, \teff, [Fe/H], \logg, and micro-turbulence $\xi$),
 respectively.  We calculate also the NLTE correction uncertainties introduced
 by the errors of these parameters.  The median values of the uncertainties are
 $\sim0.02 - 0.035$ dex from each parameter, which are negligible.  The overall
 NLTE corrections, as seen from  Fig.~\ref{fig:nlte}, are not significant for
 stars in our sample, even the largest departure is less than or equal to the
 value of the typical Li abundance error ($\sim0.2$ dex) in LTE.  Table
 ~\ref{tab:nlte} lists the LTE and NLTE results, together with other chemical
 abundances and the stellar parameters that we use in this study.  Hereafter we
 use \ali to represent \alinlte.

   \begin{sidewaystable}
          \centering
          \caption{ Stellar parameters, the selected chemical abundances, and their corresponding uncertainties of stars in our final sample.
          }
          \begin{tabular}{@{}llccccccccc@{}}
                  \hline
                  cname          &
                  object         &
                  $A_{Li}^{LTE}$ &
                  \alinlte       &
                  category       &
                  \alpfe         &
                  \feh           & 
                  \logg          &
                  \teff          &
                  [Ba II/Fe]     &
                  [Y II/Fe]      \\
                  \hline
                  18342649-2707171  & 100001      & 2.63 $\pm$ 0.21 & 2.63 & meas. & 0.05$\pm$  0.13 & 0.11 $\pm$  0.10  & 4.17 $\pm$  0.23 & 6049 $\pm$  114 & -0.09$\pm$  0.17 & -0.11$\pm$ 0.18\\
                  18253560-2525493   & 100003     & 2.64 $\pm$  0.22 & 2.61 & meas. & -0.06$\pm$  0.12 & -0.19 $\pm$  0.10  & 4.24 $\pm$  0.23 & 5958 $\pm$  112 & 0.16$\pm$ 0.17 & -0.13 $\pm$ 0.18\\
                 ...                 & ...               & ...  & ...  & ...   & ...         & ...       & ...      & ...   & ...  & ...\\
                  18225412-3410286     & 100024            & 0.49     & 0.62 & La.upper    & 0.30   $\pm$  0.14  & -0.45  $\pm$  0.11 & 4.22   $\pm$  0.3  & 4858  $\pm$  139 & -0.37  $\pm$  0.18  & 0.42   $\pm$  0.18\\
                  19241832+0057159 & 100733229         & 1.22    & 1.31 & La.upper    & 0.11   $\pm$ 0.13   &   0.13   $\pm$  0.10  & 4.38   $\pm$  0.22 & 5483  $\pm$  119 & -0.17  $\pm$ 0.18 & -0.12  $\pm$  0.17\\
                 ...                 & ...               & ...  & ...  & ...   & ...         & ...       & ...      & ...   & ...  & ...\\
                  \hline
          \end{tabular}
          \tablefoot{ Regarding Li abundance,
          both LTE values $A_{Li}^{LTE}$  from the GES iDR4 catalog and corrected NLTE results \alinlte are listed.
          The fifth column specifies to which category the star belongs to:
          "meas." means the checked Li measurements,
          and "La.upper" indicates the labeled upper limits.
          The full table is available online 
           at the CDS via anonymous ftp to cdsarc.u-strasbg.fr (\url{130.79.128.5}) or via
           \url{http://cdsarc.u-strasbg.fr/viz-bin/qcat?J/A+A/610/A38}
          } 
          \label{tab:nlte}
           \end{sidewaystable}

\begin{figure*}
            \centering
            \includegraphics[width=0.95\textwidth,angle=0]{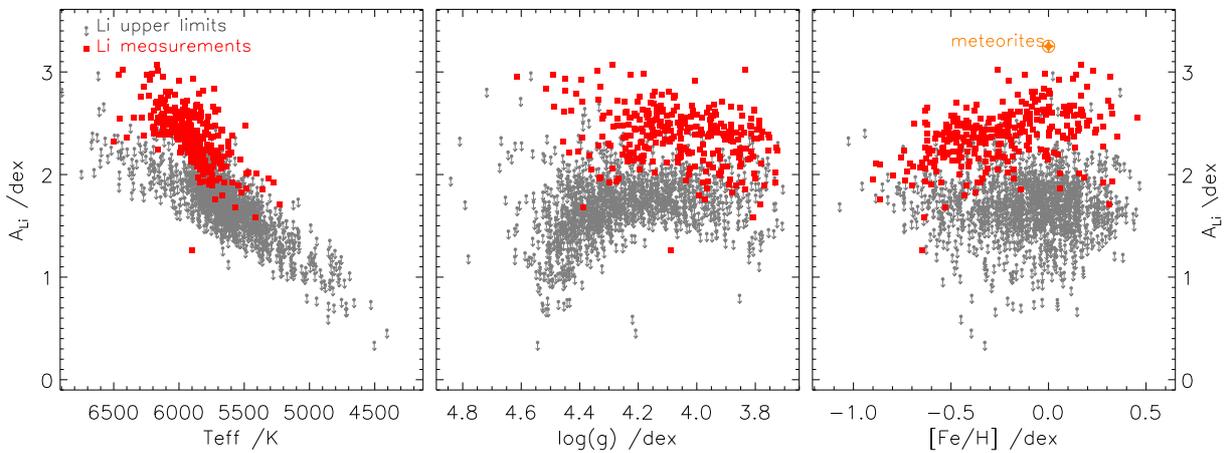}
            \caption{NLTE Li abundance for stars in our final sample as a function of 
            \teff (\textit{left}), \logg (\textit{center}), and [Fe/H] (\textit{right}), respectively.
            In all three panels,  filled red squares are the stars with checked Li measurements (302 stars), 
            and grey dots represent those labeled with Li upper limits in GES iDR4 catalog (1097 stars).
            The Li abundance of meteorites is also marked in the right panel.
            }
            \label{fig:ali}
    \end{figure*}

   In Fig.~\ref{fig:ali} we present the general behaviours of \ali against
   \teff, \logg, and [Fe/H],  respectively.  Cooler stars show lower Li
   abundances as expected (see the left panel) because they have a deeper
   surface convective zone which burns Li.  The coolest stars with Li upper
   limits are also those that have larger \logg (see the central panel).  They
   are  lower main sequence stars with small stellar  mass since \logg is an
   indicator of the evolutionary phase.  The upper envelope of the Li evolution
   with [Fe/H] (see the right panel of Fig.~\ref{fig:ali}) is traditionally
   believed to reflect the Li enrichment history of the Galactic ISM, taking
   [Fe/H] as an index of the total metallicity \citep[see ][]{Abia1998,
   romano99, romano01, travaglio01, prantzos12}.

   In the following sections we combine our Li abundances with \alpfe, \bafe,
   and \yfe ratios to shed new light on the  Li enrichment histories of
   different Galactic disc components.

 \subsection{Distinction between the Galactic thick and thin discs}
  \label{subsec:disc}

    \begin{figure*}
            \centering
            \includegraphics[width=0.95\textwidth,angle=0]{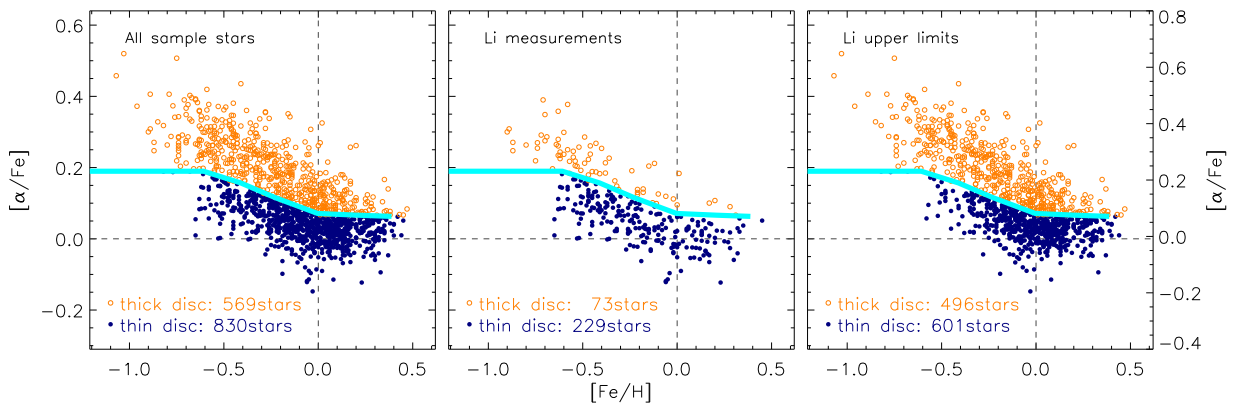}
            \caption{ Tentative separation between  thick and thin discs in the \alpfe-[Fe/H] plane
            for all sample stars (left panel), 
            stars with Li measurements (central panel),
            and stars with Li upper limits (right panel).
            In all panels the vertical and horizontal dashed lines indicate solar values.
            Filled blue dots represent the Galactic thin disc stars
            while open orange circles are the thick disc objects.
            The cyan line shows the division proposed by \citep{Adibekyan2012}.
            }
            \label{fig:disc3}
    \end{figure*}

  A precise determination of the Galactic thick and thin disc membership
  requires a detailed knowledge of the space motions, which are calculated from
  the stellar distances, proper motions and radial velocities.  Unfortunately
  only nine stars in our sample have this information from $Gaia$ DR1 TGAS
  catalog \citep{tgas2015, gaia2016}.  We are looking forward to Gaia DR2,
  which will be released in spring 2018 and will provide the five-parameter
  astrometric solutions and help us to draw a precise map of the thick and thin
  discs for almost the entire $Gaia$ catalog.

   The abundance of $\alpha$ elements relative to iron (\alpfe) are often used
   to chemically separate the Galactic thick disc from thin disc stars when the
   space motion information is lacking \citep[e.g.][]{Fuhrmann1998,
   Gratton2000, Reddy2003, Venn2004, Bensby2005, Rojas2017}.  The $\alpha$
   elements (O, Ne, Mg, Si, S, Ar, Ca, and Ti) are mostly produced by core
   collapse SNe (mainly Type II SNe) on short time scales, while iron is also
   synthesized in Type Ia SNe on longer time scales (after at least one white
   dwarf has been formed).  Stars  formed  shortly after the ISM has been
   enriched by SNe II  have enhanced \alpfe ratios, while those  formed
   sometime after  SNe Ia have contributed most of Fe  have higher iron
   abundances and lower \alpfe ratios.  Thus the  \alpfe ratio is a
   cosmic-clock, or better, it echoes the formation history \citep[see e.g.
   ][and the references therein]{Tinsley1979, Matteucci1986, Haywood2013}.  The
   Galactic thick disc stars usually have lower [Fe/H] values and higher
   \alpfe, while most of   the thin disc stars tend to have higher iron
   abundance and lower \alpfe values.  However the criterion adopted for
   separation  differs in different works.

   We define \alpfe ratios for the stars in our sample, taking into account
   four $\alpha$ elements (Mg, Ca, Si, and Ti):
  \begin{equation}
          n(\alpha)=n(\mathrm{Mg~I})+ n(\mathrm{Ca~I})+ n(\mathrm{Si~I}) + n(\mathrm{Ti~I}) + n(\mathrm{Ti~II})
      \label{equ1}     
  \end{equation}
  \begin{equation}
          $\alpfe$ = \mathrm{log}(\frac{n(\alpha)}{n(\mathrm{Fe})})_{\ast} - \mathrm{log}(\frac{n(\alpha)}{n(\mathrm{Fe})})_{\odot} 
      \label{equ2}
  \end{equation}
  The other elements (and isotopes) are not included because they are not
  measured in all our sample stars.  By taking into account the measurement
  uncertainties of the six parameters in Equation ~\ref{equ1} and ~\ref{equ2}
  (A(Mg I), A(Si), A(Ca I), A(Ti I), A(Ti II), and \feh), we derive the mean
  value of \alpfe and its corresponding 1~$\sigma$ uncertainty using the Markov
  chain Monte Carlo (MCMC) code emcee \citep{emcee2013}.  After a burn-in phase
  of 100 steps to ensure that the chains have converged, we performed 150 MCMC
  steps for each star.  The final results are listed in Table~\ref{tab:nlte}.

    Here we perform a tentative separation based on \alpfe.  We adopt a
    separation method similar to \citet{Recio-Blanco2014} and
    \citet{Mikolaitis2014} that divides the sample stars into several [Fe/H]
    \citep[\mh in the case of][]{Recio-Blanco2014, Mikolaitis2014} bins and
    finds the  possible \alpfe demarcation between the thick and thin discs in
    each interval.  Our resulting tentative division is consistent with the
    separation proposed by \citet{Adibekyan2012} who separate the high- and
    low-$\alpha$ stars with very high resolution and high S/N data.  In
    Fig.~\ref{fig:disc3} we show the separation for all sample stars, the
    division for stars with Li measurements and those with Li upper limits are
    also displayed.  There are 569 stars in our sample  possibly belonging to
    the thick disc (73 of them have Li measurements and 496 are stars with Li
    upper limits) and  830 stars that are likely thin disc members (229 stars
    are from the category of Li measurements and 601 stars have Li upper
    limits).  Our tentative separation is essentially similar to the divisions
    adopted for slightly more metal-poor stars in GES data by
    \citet{Recio-Blanco2014, Mikolaitis2014} in the \alpfe-[M/H] and
    [Mg/M]-[M/H] planes.  We choose not to adopt that because they are based on
    GIRAFFE spectra and a previous data release.

     \begin{figure}
             \centering
             \includegraphics[width=0.5\textwidth,angle=0]{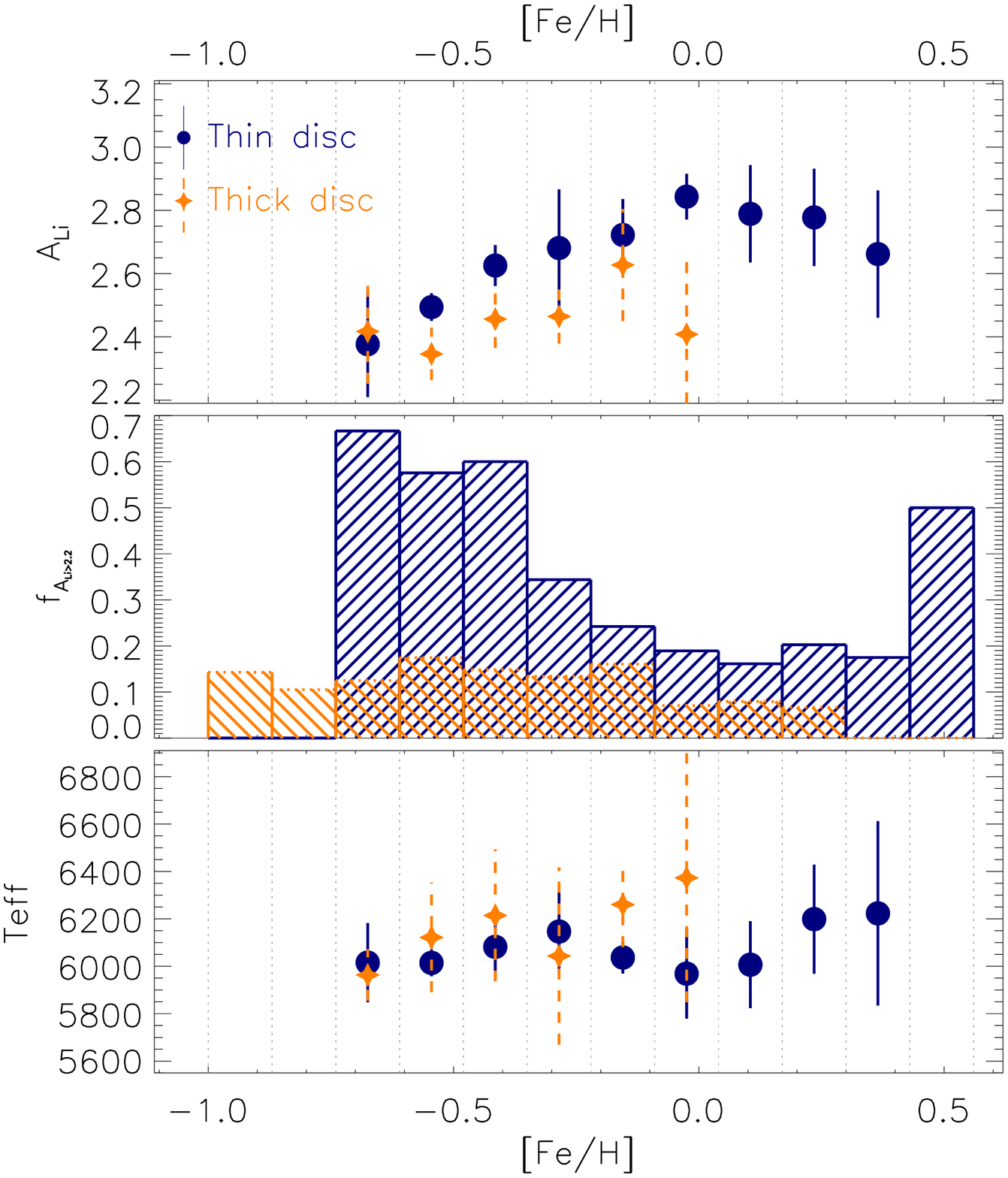}
             \caption{
                     Stars with enriched Li (\ali $> 2.2$ dex). 
             $Upper~ panel:$ 
             Error-weighted mean \ali values of the six stars with the highest Li abundance in each \feh bin,
             the standard deviation of their \ali values is considered as the uncertainty.
             Blue filled dots represent the thin disc
             and the orange stars indicate the thick disc.
             The symbols are displayed when there are more than six Li-enriched thin/thick stars in the bin. 
             $Middle~ panel:$
             Fractions of Li-enriched stars ($f_{A(Li)>2.2}$)
             in each [Fe/H] interval
             for thin disc stars (blue-shades histogram) 
             and thick disc stars (orange-shaded histogram).
             $Lower~ panel:$
             Error-weighted mean \teff values of the six stars with the highest Li abundance in each [Fe/H] bin,
             the standard deviation of their \teff values is considered as the uncertainty.
             Symbols are the same as in the upper panel.
               }
             \label{fig:li_disc}
     \end{figure}

Fig.~\ref{fig:li_disc} shows the behaviour of Li enrichment in the two discs,
when our sample stars are divided according to the criterion outlined above.
All stars with \ali $>2.2$ dex (the Spite Plateau value) are Li-enriched
compared to Pop. II stars.  Since we are not able to precisely disentangle
thick from thin disc stars in our sample without the space motion information,
we can not simply select the highest \ali as the initial Li value.  Therefore,
we adopt the method used in \citet{Lambert2004, delgadomena15, guiglion16} that
selects the six stars with the highest \ali in each \feh  bin and calculates
their mean \ali values (weighted by the reciprocal of Li abundance errors in
LTE) to track the trend of Li enrichment, the standard deviations of their \ali
is considered the corresponding uncertainty.  In the upper panel of
Fig.~\ref{fig:li_disc} the  blue filled dots and the orange stars display these
trends for the Galactic thin disc and thick disc, respectively.  Our Li trend
of thin disc stars is  systematically lower than the one proposed by
\citet{guiglion16}, while is compatible with the overall trend of
\citet{delgadomena15}.  This may be because of sample selection effects.
Strictly speaking, the trends of main sequence mean \ali is  a lower limit to
the real Li evolution in the thin and thick disc, respectively.  Some depletion
(although perhaps little) has had to happen in both groups of stars.  The
fraction of the Li-enriched stars in each \feh interval represents the overall
level of  Li enrichment.  In the middle panel of Fig.~\ref{fig:li_disc} we
present the fraction of Li-enriched stars in the thin (blue histogram) and
thick (orange histogram) discs.  The thin disc  has much higher  Li-enriched
star fractions compared to the thick disc.  We perform a K-S test to compare
the two distributions, the maximum deviation between the cumulative
distribution of the two histograms is D=0.75, and the significance level of the
K-S statistic is 0.0009.  As mentioned in Sec.~\ref{subsec:sample}, \teff is a
key parameter to trace the stellar Li destruction (stars with lower \teff have
lower \ali).  In order to investigate whether thick disc stars have experienced
a stronger Li destruction than the thin disc stars, we compare the mean \teff
(weighted by the reciprocal of \teff errors) of the six stars with highest \ali
in each \feh bin, and take their \teff standard deviation as the uncertainty,
the results are displayed in the lower panel of Fig.~\ref{fig:li_disc}.  It
becomes apparent  that the thick disc stars do not show a lower \teff compared
to the thin disc stars, in some of the \feh bins they even have higher \teff
than the thin disc ones though their \ali is lower as shown in the upper panel
of the figure.  As seen in all the three panels of Fig.~\ref{fig:li_disc}, we
conclude that the Galactic thin disc  has higher \ali and higher overall level
of Li enrichment than the thick disc, the Li trend difference between the thick
and thin discs is not due to the stellar destruction, but reflects different
initial Li abundances of the two discs.

 \subsection{Li-\alpfe anticorrelation}
   \label{subsec:lialp}

In addition to the tentative thick/thin disc separation,
we investigate the relation between \ali and \alpfe,
especially for stars with actual Li measurements,
in order to gain a deeper insight into the Li enrichment histories 
of the Galactic thick and thin disc.

     \begin{figure}
             \centering
             \includegraphics[width=0.5\textwidth,angle=0]{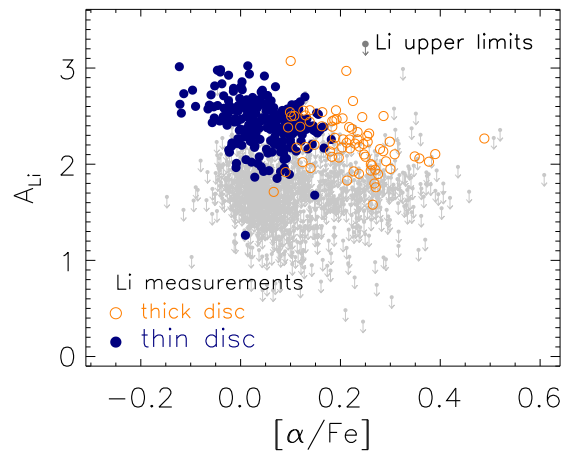}
             \caption{ NLTE Li abundance verses \alpfe. 
                     Stars with Li measurements are separated
                     as thin (filled blue dots) and thick (open orange circles) disc stars,
                     as the same as in the middle panel of  Fig.~\ref{fig:disc3}.
                     The grey dots represent sample stars with  Li upper limits.
                      }
             \label{fig:alialp}
     \end{figure}

\begin{figure*}
             \centering
             \includegraphics[width=0.85\textwidth,angle=0]{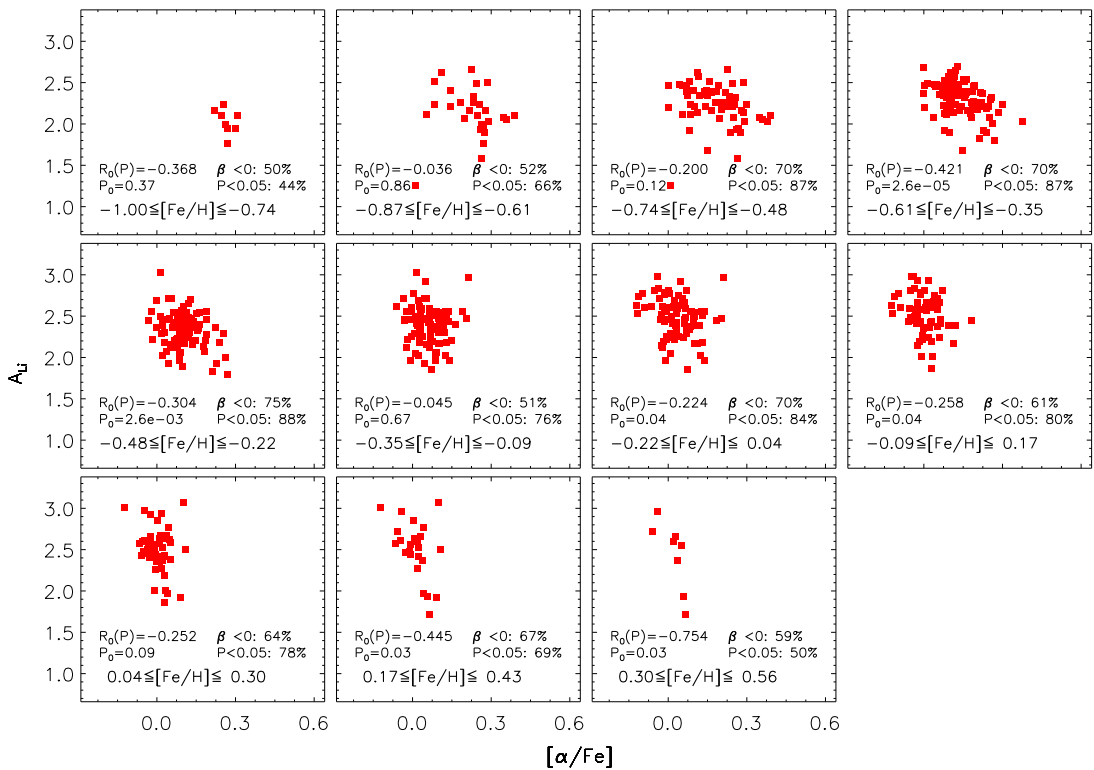}
             \caption{
                     NLTE Li abundance as a function of \alpfe in each [Fe/H] bin for our sample stars with Li measurements. 
             The range of [Fe/H] is specified in each plot.
             The [Fe/H] bin size is 0.26 dex.
             In each bin 
             the mean correlation coefficient ($R_0(P)$) between \ali and \alpfe,
             the mean correlation significance level ($P_0$),
             the possibility of \lialpha ($\beta$<0) if consider the abundance uncertainties,
             and the possibility of P<0.05 for the anticorrelation
             are labeled.          }         
             \label{fig:binalialp}
     \end{figure*}

  In Fig~\ref{fig:alialp} we display the overall relation between \ali and
  \alpfe.  A Li-\alpfe anticorrelation is clearly seen in this figure and is
  highly statistically significant (its Pearson's correlation for stars with Li
  measurements has confidence level > 99\%).  However one can not ignore that
  as \feh increases, \ali rises (see the right panel of Fig.~\ref{fig:ali})
  while \alpfe decreases (see Fig.~\ref{fig:disc3}).  To eliminate the \feh
  evolutionary effect in the Li-\alpfe anticorrelation, one should compare
  stars with similar \feh values.  For this purpose we present \ali as a
  function of \alpfe in  different [Fe/H] bins in Fig~\ref{fig:binalialp}.  The
  bin size  is 0.26 dex, which is twice  the maximum error on [Fe/H] for our
  sample stars.  To avoid the bin selection bias,  0.13 dex is overlapped
  between the adjacent [Fe/H] bins.  We calculate the mean values of Pearson's
  correlation coefficient ($R_0(P)$) and its significance level $P_0$ between
  \ali and \alpfe in each \feh bin.  We also derive the possibility of the
  anticorrelation in each bin taking into account the abundance uncertainties.
  In this case, Li abundance error in LTE measurement is considered as the
  uncertainty of NLTE results, and a Bayesian linear regression method
  described in \citet{Kelly2007} is performed.  If the number of stars in the
  bin is greater than 10, then the Markov chains will be created using the
  Gibbs sampler, otherwise the Metropolis-Hastings algorithm is used.  Negative
  value of the linear regression slope  $\beta$ means anticorrelation.  For
  each \feh bin, we calculate the possibility of an anticorrelation ($\beta$<0)
  of Li-\alpfe, and find that the possibility is $\gtrsim 68\%~(1\sigma)$ for
  half of them, with a corresponding $1-2\sigma$ significance of P < 0.05 (i.e.
  the confidence interval is greater than 95\%).  This means that, at similar
  \feh, \ali is very likely anti correlated with \alpfe for field stars.

\begin{figure*}
             \centering
             \includegraphics[width=0.85\textwidth,angle=0]{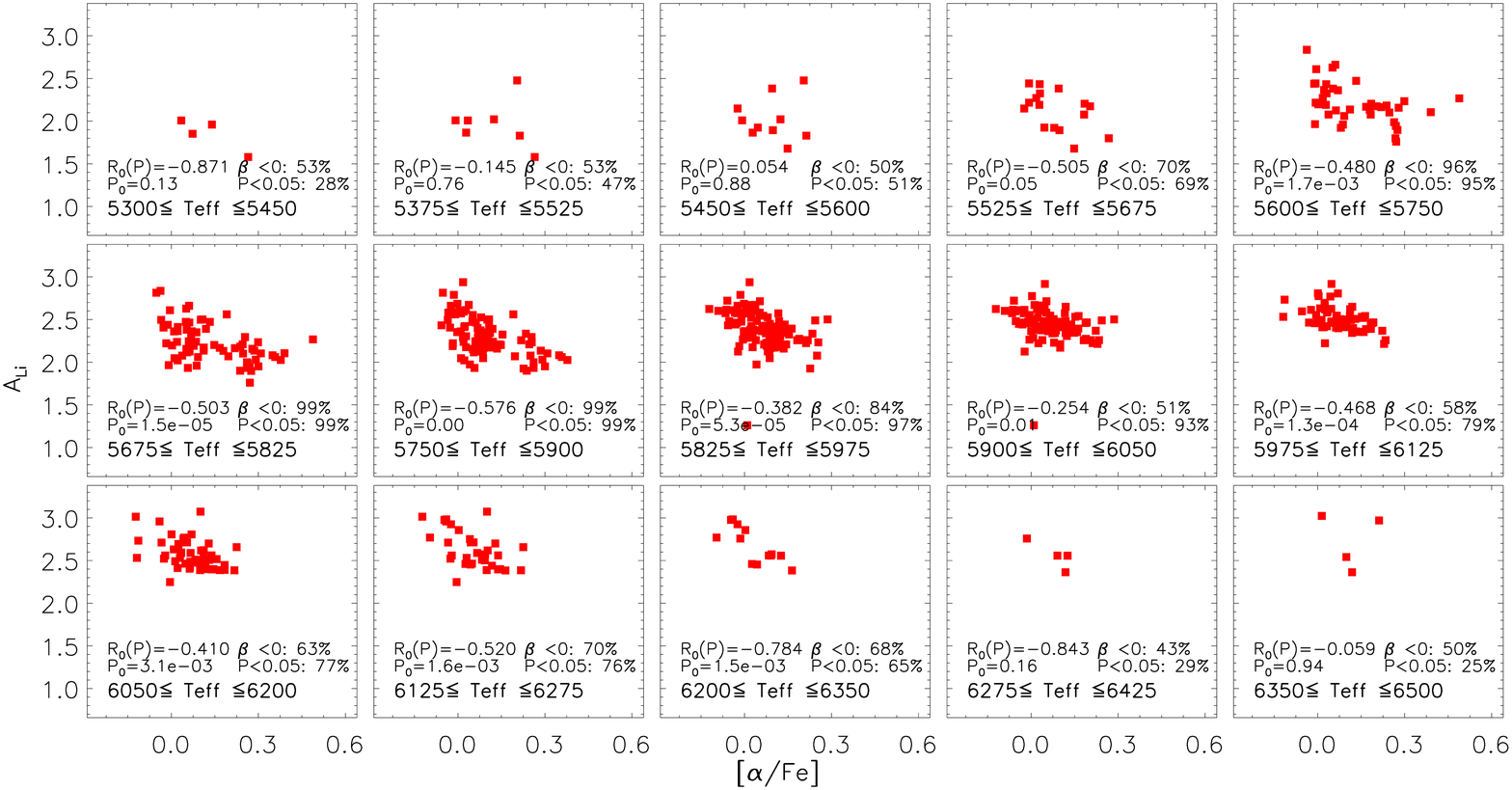}
             \caption{
                     NLTE Li abundance as a function of \alpfe in each \teff  bin for our sample stars with Li measurements. 
             The range of \teff is specified in each plot, with bin size of 150 K.
             The mean correlation coefficient ($R_0(P)$) between \ali and \alpfe,
             the mean correlation significance level ($P_0$),
             the possibility of \lialpha ($\beta$<0) if consider the abundance uncertainties,
             and the possibility of P<0.05 for the anticorrelation
             are labeled.          }         
             \label{fig:binaliteff}
     \end{figure*}

  We have shown in Fig.~\ref{fig:ali} that  \ali decreases with lower \teff, as
  \teff is a tracer of the Li destruction.  In Fig.~\ref{fig:binaliteff} we
  show it is highly probable that stars with similar \teff have their Li
  abundances anti-correlated with \alpfe.  Since 98\% of our sample stars have
  \teff error less than 150 K, we choose this value as the bin size in
  Fig.~\ref{fig:binaliteff}.  To avoid the bin selection bias we overlap 75 K
  between the adjacent \teff bins.  The anticorrelation between the mean values
  of \ali and \alpfe is significant at the level of P$\le$0.05, except for the
  intervals with too few stars (the first three and last two bins).  If the
  abundance uncertainties are considered, we use the same Bayesian linear
  regression method as in   Fig~\ref{fig:alialp} to derive the \lialpha
  possibility and its corresponding significance level, the results are listed
  in Fig.~\ref{fig:binaliteff}.

\begin{figure*}
             \centering
             \includegraphics[width=0.85\textwidth,angle=0]{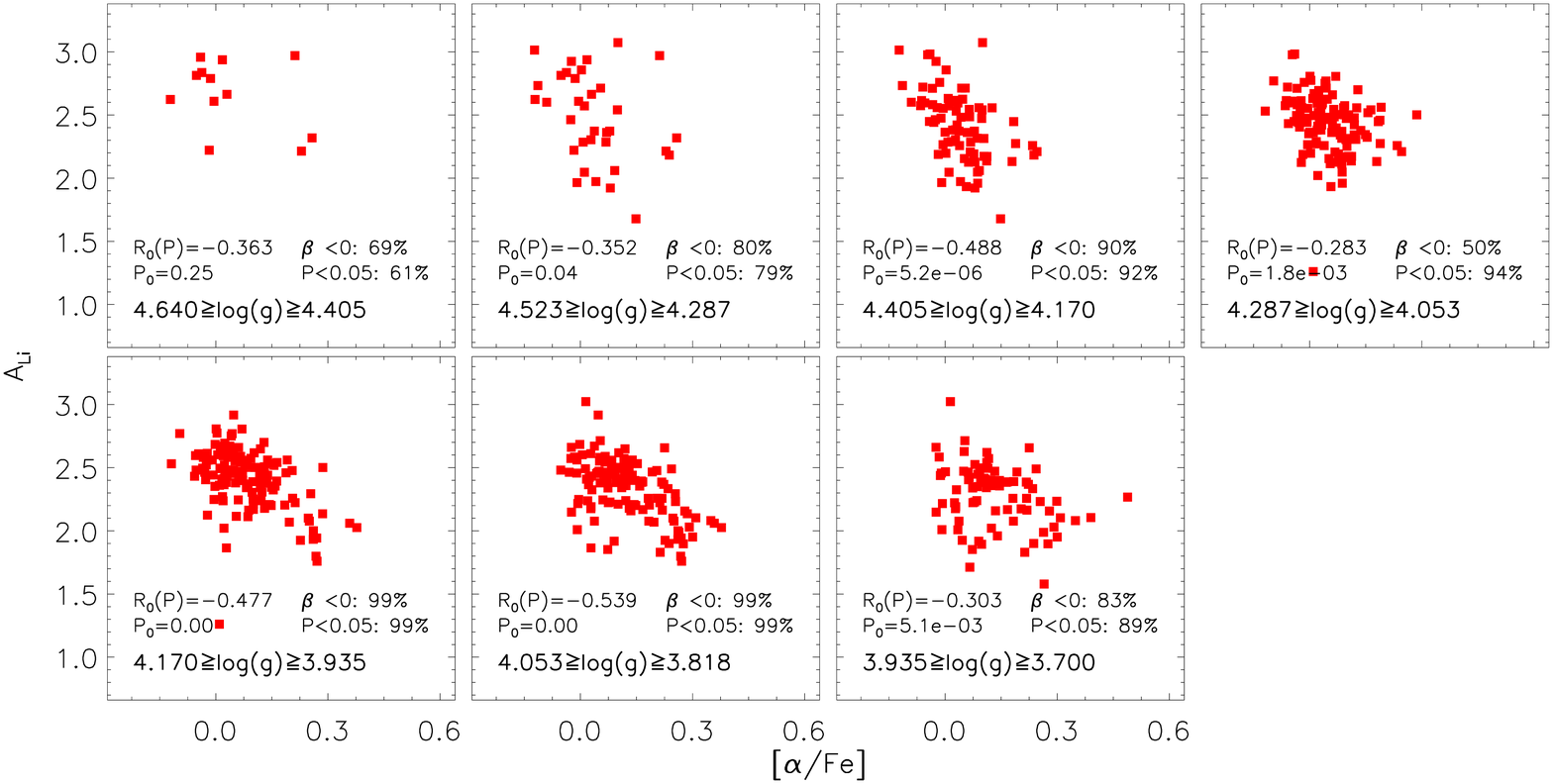}
             \caption{
                     NLTE Li abundance as a function of \alpfe in each \logg  bin for our sample stars with Li measurements. 
             The range of \logg is specified in each plot, with bin size of 0.235 dex.
             The mean correlation coefficient ($R_0(P)$) between \ali and \alpfe,
             the mean correlation significance level ($P_0$),
             the possibility of \lialpha ($\beta$<0) if consider the abundance uncertainties,
             and the possibility of P<0.05 for the anticorrelation
             are labeled.          }         
             \label{fig:binalilogg}%
     \end{figure*}

 It is of great interest to compare the correlation between \ali and \alpfe for
 stars with similar evolutionary phase (position in the main sequence) as well.
 We use \logg as an index of the evolutionary phase and present the Li-\alpfe
 anticorrelation in Fig~\ref{fig:binalilogg}.  The \logg bin size is 0.235 dex,
 and is based on a criterion on the \logg error that covers 68\% of the sample
 stars.  Similar to our treatment for \teff and \feh, we overlap half of the
 bin size  between the adjacent bins to avoid selection bias.  In each \logg
 interval the mean \ali-\alpfe correlation is negative, and the mean
 correlation significance levels ($P_0$) are less than 0.05 except in the first
 bin which has only a few stars.  When taking into account the abundance
 uncertainties as described above for different \teff and \feh bins, there is a
 (very) high possibility  for the \lialpha in almost every \logg bin.

\begin{figure*}
             \centering
             \includegraphics[width=0.8\textwidth,angle=0]{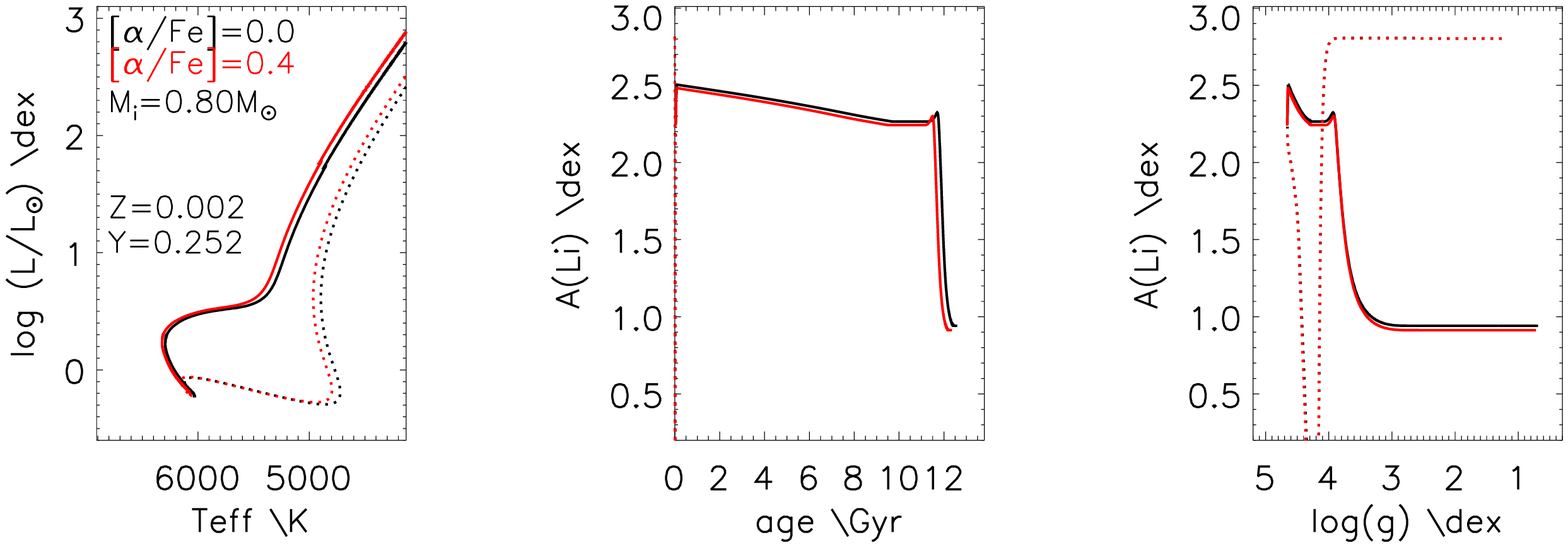}
             \caption{Li evolution from the stellar evolution model \parsec.
             Two stars with the same metallicity and He content (Z=0.002, Y=0.252), 
             the same initial Li abundance (\ali=2.8 dex), 
             the stellar mass (M$_i$=0.80 \msun), 
             but different $\alpha$-enhancements are compared.
             The red curves indicate the star with \alpfe=0.4 dex
             while the black ones are for the star with \alpfe=0 (solar-scaled).
             The dotted lines in both colors represent the evolution before the zero-age main sequence (ZAMS),
             and the solid lines indicate the evolution after.
             The panels from left to right are
             HR diagram, \ali evolution with  stellar age, 
             and \ali as a function of \logg, respectively.} 
             \label{fig:limodel}
     \end{figure*}

      To ensure that the \lialpha echoes the initial \ali and is not affected
      by any  stellar evolution effect, we check the behaviour of A(Li) in
      stellar evolution models.  We use the latest version of \parsec
      \citep{bressan2012, fu2017} to calculate Li evolution for two stars with
      the same metallicity, helium content, stellar mass, initial Li abundance,
      but one having  \alpfe=0.4 dex and the other  \alpfe=0 (solar-scaled).
      Though the star with \alpfe=0.4 dex is hotter in every evolutionary phase
      (see the left panel of Fig.~\ref{fig:limodel}), there is no measurable
      difference between its \ali and that of the star with \alpfe=0, neither
      at the same age (see the central panel) nor at the same \logg.  The
      different \ali behaviour in stars with different \alpfe has to come from
      their initial Li abundances which reflect different Li  enrichment
      histories in different disc components.

 The \lialpha also  sheds light on the role of core collapse SNe in the total Li enrichment.
 Since core collapse SNe are the main producer of the $\alpha$ elements,
 if these stars are responsible also for a major Li production, 
 this element should increase with \alpfe.
 Our Li-\alpfe anticorrelation result, instead, implies that 
 the contribution from core collapse SNe to the global Li enrichment is unimportant, if not negligible,
  thus supporting previous findings (see the Introduction).

  \subsection{Li-$s$(-$process~elements$) correlation}
  \label{subsec:lis}

\begin{figure*}
             \centering
             \includegraphics[width=0.95\textwidth,angle=0]{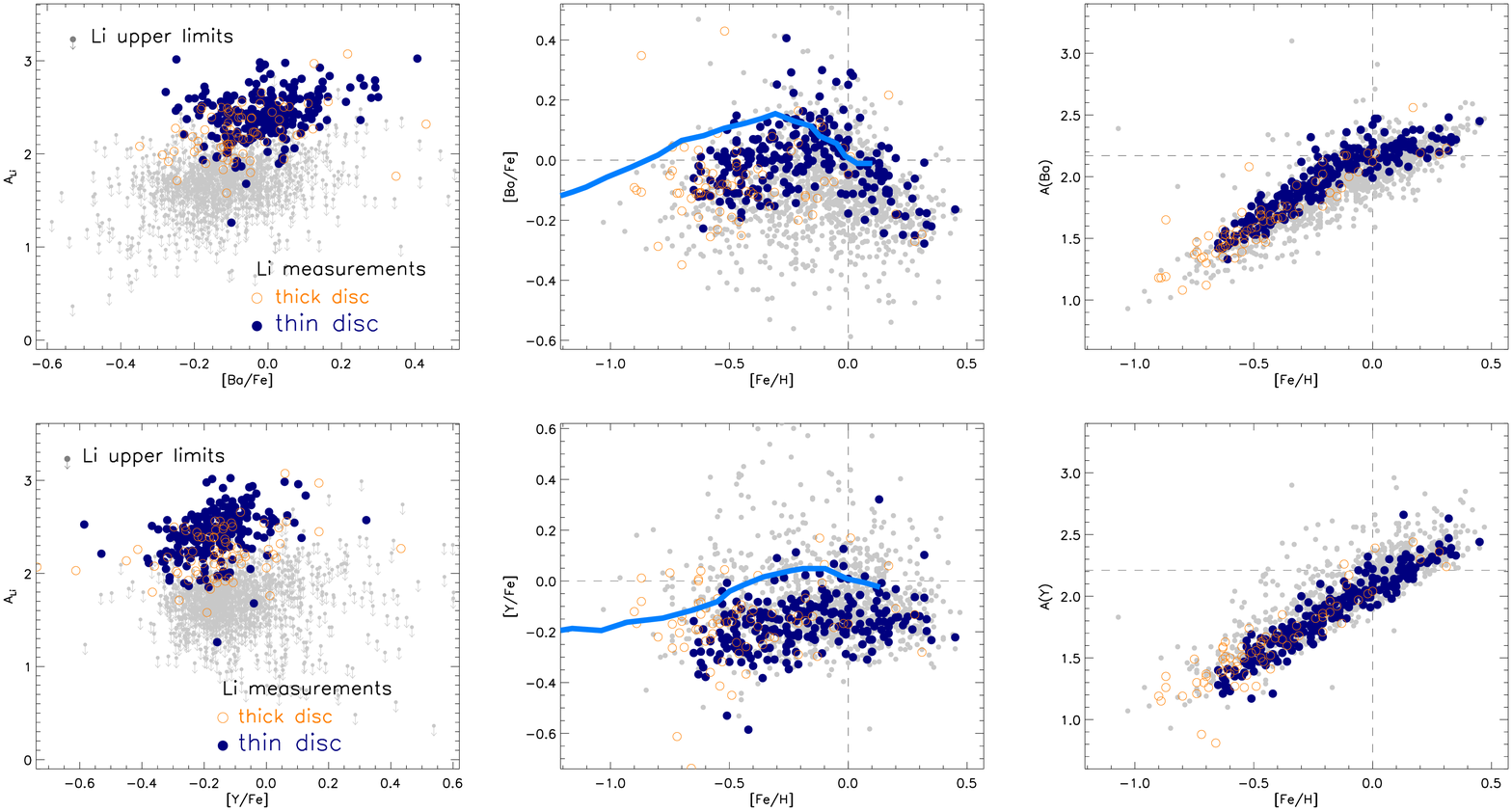}
             \caption{
                     NLTE Li abundance versus \bafe  (upper left panel),
             \bafe against \feh (upper middle panel),
             and absolute barium abundance A(Ba) as a function of \feh (upper right panel).
             The three lower panels are the same for yttrium. 
             The \bafe and \yfe versus [Fe/H] trends predicted by \citet{Bisterzo2017} for the Galactic thin disc
             are overlaid with light blue solid lines.
             Dashed lines indicate the solar values.
             Stars with Li measurements are separated as thick (filled blue dots) 
             and thin (open orange circles) disc stars,
             as the same as in the middle panel of  Fig.~\ref{fig:disc3}.
             The grey dots represent sample stars with  Li upper limits.
             }  
             \label{fig:bafeh}
     \end{figure*}

  Similarly to the \alpfe ratio, the ratio of the slow ($s-$) neutron capture
  process elements to iron can be regarded as a  cosmic clock.  Ba, Sr , La,
  and Y are mainly $s-$process elements produced on long timescales by low mass
  AGB stars \citep{Matteuccibook}.  Since a low mass star must evolve to the
  AGB phase before the $s-$process can occur, the $s-$process  elements are
  characterized by a delay in the production, much like the delay of iron
  production by SNe Ia relative to the $\alpha$ elements production by core
  collapse SNe.  Among the four $s-$process elements mentioned above, GES
  provides the abundances of Y II (the first $s-$process peak element) and Ba
  II (the second $s-$process peak element) for all our sample stars.  Their
  abundances behave differently in the Galactic thick and thin discs
  \citep{Bensby2005, israelian2014, bensby2014, Bisterzo2017, mena2017}.
  Unlike the Galactic thick disc stars, which show an almost constant [Ba/Fe]
  abundance close to the solar value, the Galactic thin disc stars have their
  [Ba/Fe] abundances increasing with [Fe/H] and reaching their maximum values
  around solar metallicity, after which a clear decline is seen \citep[see also
  ][for the most recent $s-$process calculation in AGB yields]{Cristallo2015b,
  Cristallo2015}.  The same trend is observed in our sample.  In
  Fig.~\ref{fig:bafeh} we display the Li-[Ba/Fe], \bafe as a function of
  [Fe/H], and the evolution of  absolute Ba abundance A(Ba), as derived from Ba
  II lines.  The same figures are also plotted for yttrium (Y II).  \bafe and
  \yfe values here are derived from MCMC simulations, taking into account the
  measurement uncertainties of A(Ba II)/A(Y II) and \feh.  By applying the same
  MCMC setups used for \alpfe (see Sec. ~\ref{subsec:disc}), we calculate the
  mean values of \bafe and \yfe for each star.  These values, together with
  their corresponding 1 $\sigma$ uncertainties, are listed in Table
  \ref{tab:nlte}.  In the literature there are several theoretical works on the
  evolution of \bafe and \yfe  in the  Galactic thin disc \citep[e.g.
  ][]{Pagel1997, Travaglio1999, Travaglio2004, Cescutti2006,  Maiorca2012,
  Bisterzo2017}.  For comparison, we show  in Fig.~\ref{fig:bafeh} the
  predictions of the most recent one \citep{Bisterzo2017} who use the updated
  nuclear reaction network.

\begin{figure*}
             \centering
             \includegraphics[width=0.85\textwidth,angle=0]{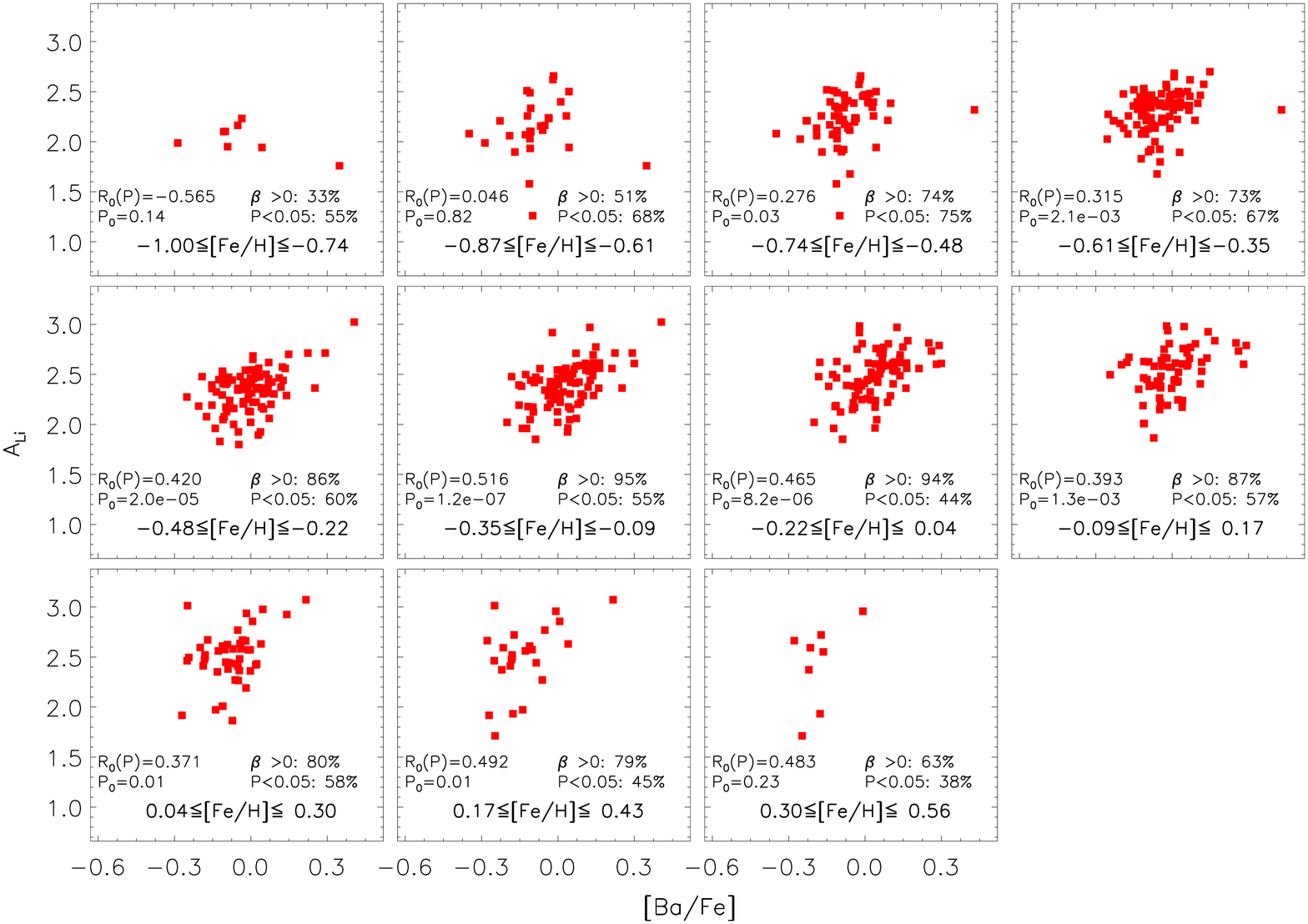}
             \caption{
                     NLTE Li abundance as a function of \bafe in each \feh bin 
                     for our sample stars with Li measurements. 
             The range of [Fe/H] is specified in each plot.
             The [Fe/H] bin size is 0.26 dex.
             In each bin 
             the mean correlation coefficient ($R_0(P)$) between \ali and \bafe,
             the mean correlation significance level ($P_0$),
             the possibility of \ali-\bafe correlation ($\beta$>0) if consider the abundance uncertainties,
             and the possibility of P<0.05 for the anticorrelation
             are labeled.          }         
             \label{fig:binaliba}
     \end{figure*}

\begin{figure*}
            \centering
            \includegraphics[width=0.85\textwidth,angle=0]{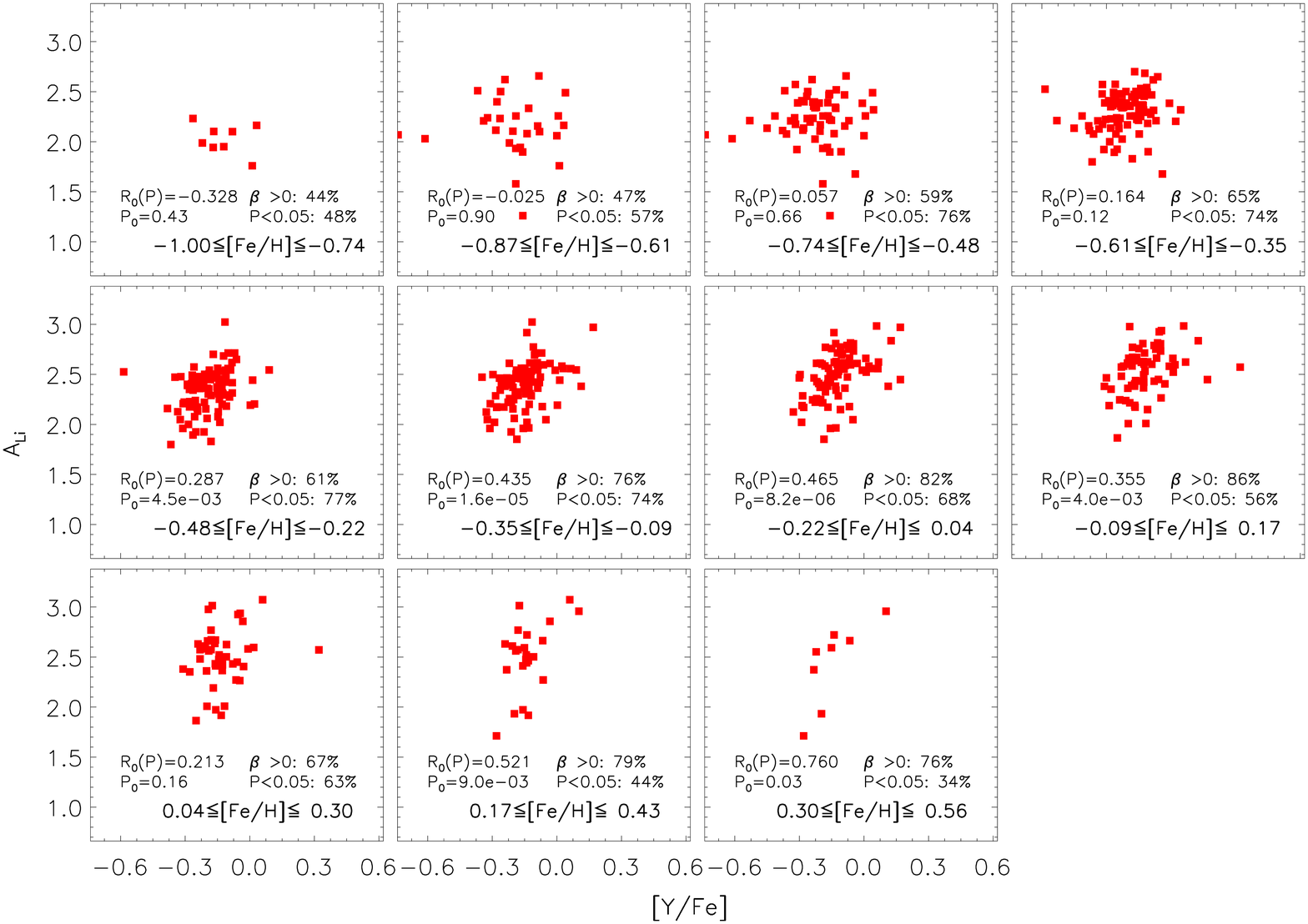}
             \caption{
                     NLTE Li abundance as a function of \yfe in each \feh bin 
                     for our sample stars with Li measurements. 
             The range of [Fe/H] is specified in each plot.
             The [Fe/H] bin size is 0.26 dex.
             In each bin 
             the mean correlation coefficient ($R_0(P)$) between \ali and \yfe,
             the mean correlation significance level ($P_0$),
             the possibility of \ali-\yfe correlation ($\beta$>0) if consider the abundance uncertainties,
             and the possibility of P<0.05 for the anticorrelation
             are labeled.          }         
            \label{fig:binaliy}
    \end{figure*}

  For stars with true Li measurements (filled blue dots and open orange circles in Fig.~\ref{fig:bafeh}),
  their \ali show a significant overall correlation with \bafe and \yfe (at a 99\% confidence level).  
  Fig~\ref{fig:binaliba} and  Fig~\ref{fig:binaliy} display
  the \ali-\bafe and \ali-\yfe correlation, respectively, in different [Fe/H] intervals. 
  The anticorrelation is especially significant 
  in the range -0.74$\lesssim$\feh$\lesssim$0.43 for \ali-\bafe
  and -0.48$\lesssim$\feh$\lesssim$0.56 for \ali-\yfe.
  The Li-$s$(-process elements) 
  correlation is easily understood since both of them 
  are produced mostly by long-lived stellar sources 
  \citep[see e.g.] [ and references therein] {romano99,romano01, Bisterzo2017}.

 The main source for $s$-process elements appear to be low mass AGB stars \citep{Kippenhahnbook}.
 Given the small [Ba/Fe] enhancement in most of the stars with lower Li 
 which have higher \alpfe and probably belong to the thick disc,
 we conclude that AGB stars may mainly contribute to the chemical enrichment in the Galactic thin disc (with lower \alpfe).
 Similar results have also been reported by \citet{Mashonkina2001, mena2017}
 and agree with existing theoretical scenarios of relatively fast/slow thick/thin disc formation 
 \citep[see e.g.][] {micali2013}.

\section{Discussion}
 \label{sec:dis}

 We show in Sec.~\ref{sec:res} that 
 the Galactic thin disc stars have stronger and higher overall level of  Li enrichment
 than the thick disc stars.
 This result gives us  new insights into other observational phenomena, 
 e.g. the "Li-rich giant problem" (see the introduction)
 which should be studied separately in different environments for two reasons.
 First, the definition of "Li-rich giant" will be affected 
 since the thick and thin disc stars have different initial abundances of Li;
 second and more important,
 giant stars in the environment with stronger Li enrichment
 are more likely to be polluted by external sources
 \citep[i.e. contamination by the ejecta of nearby novae,][]{Martin1994}.
 \citet{Aguilera2016} calculate red giant models with substellar object engulfment
 and conclude that 
 a Li enrichment up to A(Li)$\sim$2.2 dex can be explained by
 engulfing a 15 Jupiter mass ($M_J$) substellar object. 
 If we make a simple calculation, 
 a 15 $M_J$ substellar object with the initial solar metal mixture contains 
 $M_{Li}\sim$ 2 $\times 10^{-10}$ \msun,
 and a single nova outburst can eject 
 $M_{Li}$=0.3$\sim$4.8 $\times 10^{-10}$\msun ~\citep[nova V1369 Cen, see][]{izzo15}
 or $M_{Li} \sim$ 7 $\times 10^{-9}$ \msun ~\citep[nova V5668 Sgr, see][]{molaro16}.
 A contamination containing 1/30 $\sim$ 7 single nova ejecta 
 provides the same amount of Li as a 15 $M_J$ substellar object offers.
 Furthermore,
 taking into account that
 {\it i.} the nova ejecta do not increase the stellar angular momentum 
 that introduces mixing and Li burning,
 and {\it ii.} usually a white dwarf can generate multiple novea over time,
 Li enrichment introduced by novae ejecta contamination is more significant than the substellar objects engulfment. 
 So we expect different frequencies  of  Li-rich giant stars in different environments.

 The upper envelope of the \ali measurements traced by our sample stars 
 clearly shows that the Galactic Li abundance increases as [Fe/H] approaches the solar value, 
 and then declines for super-solar metallicity objects 
 (see Fig.~\ref{fig:ali}, right panel, and Fig.~\ref{fig:li_disc}, upper panel). 
 A similar trend is seen also in \citet{ramirez12, delgadomena15, guiglion16}. 
 Apparently, these findings contrast with classic Galactic chemical evolution model predictions 
 of a monotonic increase of \ali with metallicity in the Galactic disc 
 \citep[see][]{romano99, romano01, travaglio01, prantzos12}.
 Though metal-rich stars have deeper surface convective zone compared to the metal-poor ones with the same stellar mass,
 it is not likely that 
 the Li decline  is due to a stronger Li destruction during the main sequence in the super metal-rich stars 
 because the mean \teff of these stars are comparable or higher than the solar metallicity ones
 (see the lower panel of Fig.~\ref{fig:li_disc} ).
 From the stellar physics point of view
 if the main sequence destruction is not responsible for the Li decline as explained above,
 the answer may lie in two fields:
 {\it i.} special conditions of mixing and diffusion;
 and {\it ii.} depletion during the pre-main sequence.
 For the first one, \citet{Xiong2009} present a non-local convection stellar model 
 that considers a strong overshooting 
 (the bottom boundary is set at a depth deeper than $5\times10^6$ K, which can burn Li efficiently)
 and pure gravitational settling of microdiffusion,
 they predict that A(Li) of warm stars (\teff $>$ 6000K) decreases with increasing \teff
 so stars with high \teff can also have low A(Li).
 For the second one,
 \citet{fu15} discuss that 
 Li in pre-main sequence stars
 is first depleted by convection then restored
 by the residual mass accretion,
 since the very metal-rich disk has high opacity and is easily evaporated, 
 it is possible that the accretion is terminated early and the Li is not fully restored.
 In the following, we discuss some possible explanations 
 from the perspective of the Galactic chemical evolution
 for the observed Li decrease in super-solar metallicity stars, 
 in the framework of the model by \citet{romano99, romano01}.

 In particular, from a close inspection of figures 3 and 4 of \citet{romano01}, 
 we note that the Li abundance increases with metallicity until [Fe/H] $\sim$ 0, 
 but then decrease for [Fe/H] $>$ 0, in agreement with the trend suggested by our observations.
 A similar prediction trend is also seen in figure 5 of \citet{Abia1998} with a different choice of yields.
 The predicted behaviour in \citet[][with contributions from both AGB stars and core-collapse SNe]{romano01}
 is due to the presence of a threshold in the gas density in their model, 
 below which the star formation stops 
 \citep[see also ][and reference therein]{Chiappini2001}.
 During the last $\sim5$ Gyr of evolution, because of the presence of such a threshold, 
 the star formation in the solar neighbourhood
 has several short periods of activity intercut with star formation gaps. 
 As a consequence, the restitution rate of Li from AGB stars and core-collapse SNe, 
 which tracks rather closely the star formation rate owing to the relatively short lifetimes 
 of the stellar progenitors, is considerably reduced, 
 and the stellar Li astration eventually overcomes the Li production.
 This star formation gap explanation may hold as well for pure $s$-process elements
 synthesised in low mass stars, 
 and explains why their abundance relative to iron decreases in super-solar metallicity stars 
 \citep[Fig.~\ref{fig:bafeh}, the two middle panels; see also][]{Bensby2005,Bisterzo2017, mena2017}.
 Furthermore, the absolute abundances of barium show kind of a plateau at super-solar metallicity
 (see the upper left panel of Fig.~\ref{fig:bafeh}),
 which strongly indicates that the production of the $s$-process elements was minor, 
 or even stopped for some time. 
 However,  
   though the NLTE correction has little influence on Ba abundances \citep{Mishenina2015},
 stellar activities might affect the A(Ba) trend. 
   \citet{Reddy2017} present a strong correlation between the [Ba II/Fe] values
   and the chromospheric activity for solar-twin stars,
   they demonstrate that the Ba II abundance of these stars are overestimated
   because 
   of the adoption of a too low value of microturbulence in the spectrum synthesis.
 Unfortunately, we are  not able to confirm this effect in our sample stars
 because the stellar activity information is not available.
 It should also be stressed that the yields of super-solar metallicity stars 
 are not widely investigated in the literature,
 the yields of $s-$process elements are not very clear.
 For instance, 
 \citet{Cristallo2015b} present that the yields of the second $s-$process peak (including Ba)
 mainly contributes to -0.7<\feh<-0.3
 and decreases its contribution at higher metallicities,
 while \citet{Karakas2016} show that the yields of Ba increase from Z=0.007 to Z=0.03.
 Further complicating matters,
 one have to consider also the contribution of
  low-mass stars that die when the metallicity of the ISM is super-solar,
  but were born at earlier stages from a metal-poor ISM.
  Such work requires thorough calculations in the Galactic chemical models.
 
 Still, precise measurements of Li and $s$-process element abundances 
 in stars will be extremely useful to constrain the AGB yields,
 as well as the Galactic chemical evolution models.

 As recently confirmed by the direct detection of Li in the spectra of nova V1369 Cen by \citet{izzo15}, 
 given the current estimates of the Galactic nova rate,
 novae might contribute most of the Li in the Galaxy. 
 In particular slow novae, with a rate of around 17 events/yr, 
 would eject in the ISM Li amounts significantly larger than their fast counterparts 
 \citep{izzo15}. 
 The nova rate is highly sensitive to the assumed fraction and mass distribution of the binary stellar progenitors. 
 In particular, \citet{Gao2014, Yuan2015, gao2017} report that the fraction of close binaries 
 (which lead to nova outbursts)
 decreases with increasing [Fe/H]. 
 The super-Chandrasekhar SNe Ia, an end-result of binary stars, 
 also strongly prefer metal-poor environments \citep{Khan2011}. 
 These observations imply that the occurrence of nova systems may be lower at higher metallicities, 
 which would lower the total Li production from such objects in high-metallicity environments.
 Such a metal-dependence of the nova system formation rate has not been taken into account 
 in Galactic chemical evolution models up to now \citep{romano01, travaglio01}.
 Here we suggest that it could be (co-)responsible for the observed decreasing trend of \ali for [Fe/H] $>$ 0, 
 a hypothesis that needs to be tested by means of detailed chemical evolution models.

\section{Summary}
\label{sec:con}

We investigate the Li enrichment histories in the Galactic discs using GES iDR4 data.
Li abundance, \alpfe, \bafe, and \yfe for 1399
well-measured ([Fe/H] error <0.13 dex) main sequence (\logg $\le$ 3.7 dex) 
field stars with UVES spectra are studied.
We divide the sample stars into two categories:
Li measurements and Li upper limits.
NLTE corrections are applied to the Li line at 6708 \AA.
Four $\alpha$ elements, Mg, Si, Ca, Ti, are considered 
in the MCMC calculations to derive the total \alpfe and the corresponding 1 $\sigma$ uncertainty.
We find a Li-\alpfe anticorrelation independent of [Fe/H], \teff, and \logg in our sample stars with actual Li measurements.
After checking the behaviour of surface Li  in stellar evolution models
we conclude that different $\alpha$ enhancements do not lead to measurable \ali differences,
thus the Li-\alpfe anticorrelation echos different levels of  Li enrichments in these stars.
A Li-$s(-process$ elements) correlation, 
which is connected to the nucleosynthesis in their common production site (AGB stars), is also seen.

We  perform a tentative division  based on \alpfe and [Fe/H] 
in order to separate our sample into thick disc stars and thin disc ones.
  By comparing in each [Fe/H] bin the 
  error-weighted mean
  \ali and \teff values of stars with the highest Li abundance,
  as well as the fractions of Li-enriched stars, 
  we conclude that the thin disc stars 
  experience 
  a stronger Li enrichment throughout their evolution, compared to the thick disc stars.

The Li decline in super-solar metallicity, 
which has already been reported by \citet{delgadomena15} and \citet{guiglion16}, is confirmed by the GES analysis.
We discuss the possible explanations of this scenario.
It may be a joint consequence of AGB yield evolution and the low binary fraction at a high metallicity.

\begin{acknowledgements}

        This work is based on data products from observations made with 
        ESO Telescopes at the La Silla Paranal Observatory under programme ID 188.B-3002 and 193.B-0936. 
        These data products have been processed by the Cambridge Astronomy Survey Unit (CASU) 
        at the Institute of Astronomy, University of Cambridge, 
        and by the FLAMES/UVES reduction team at INAF/Osservatorio Astrofisico di Arcetri. 
        These data have been obtained from the Gaia-ESO Survey Data Archive, 
        prepared and hosted by the Wide Field Astronomy Unit, 
        Institute for Astronomy, University of Edinburgh, 
        which is funded by the UK Science and Technology Facilities Council.
        This work was partly supported by the European Union FP7 programme through ERC grant number 320360 
        and by the Leverhulme Trust through grant RPG-2012-541.  
        We acknowledge the support from INAF and Ministero dell' Istruzione, dell' Universit\`a' e della Ricerca (MIUR) 
        in the form of the grant "Premiale VLT 2012". 
        The results presented here benefit from discussions held during the
        Gaia-ESO workshops and conferences supported by the ESF (European Science Foundation) 
        through the GREAT Research Network Programme.
        This research has made use of the TOPCAT catalogue handling and plotting tool \citep{topcat,topcat2017}; 
        of the Simbad database and the VizieR catalog access tool, CDS, Strasbourg, France \citep{Ochsenbein2000}; 
        and of NASA’s Astrophysics Data System.
        X.F acknowledges helpful discussions with Nikos Prantzos and Paolo Molaro,
        and thanks Zhiyu Zhang for the help on MCMC calculations.
        E.D.M. and S.G.S. acknowledge the support from Funda\c{c}\"{a}o para a Ci\^{e}ncia e a Tecnologia (FCT)
        through national funds and from FEDER through COMPETE2020 by the following grants:
        UID/FIS/04434/2013 \& POCI--01--0145-FEDER--007672, PTDC/FIS-AST/1526/2014 \& POCI--01--0145-FEDER--016886,
        and PTDC/FIS-AST/7073/2014 \& POCI-01-0145-FEDER-016880.
        E.D.M. and S.G.S. also acknowledge the support from FCT through Investigador FCT contracts
        IF/00849/2015/CP1273/CT003  and IF/00028/2014/CP1215/CT0002.
        C.A. acknowledges to the Spanish grant AYA2015-63588-P
        within the European Founds for Regional Development (FEDER).
        A.K. and T.B. acknowledge the project grant ``The New Milky Way’’ from the Knut and Alice Wallenberg Foundation.
        R.S. acknowledges support from the Polish Ministry of Science and Higher Education.

\end{acknowledgements}

   \bibliographystyle{aa} 
   \bibliography{li_evo.bib} 

\end{document}